\documentclass[fleqn,usenatbib]{mnras}

\usepackage{newtxtext,newtxmath}

\usepackage[T1]{fontenc}

\DeclareRobustCommand{\VAN}[3]{#2}
\let\VANthebibliography\thebibliography
\def\thebibliography{\DeclareRobustCommand{\VAN}[3]{##3}\VANthebibliography}

\usepackage{graphicx}	
\usepackage{amsmath}	

\title[Ice origins of OCS]{Ice origins of OCS and chemistry of CS$_2$-bearing ice mantles}

\author[R. Mart\'in-Dom\'enech et al.]{
Rafael Mart\'in-Dom\'enech,$^{1}$\thanks{E-mail: rmartin@cab.inta-csic.es} \thanks{The reported experiments were performed during an internship at the Center for Astrophysics $|$ Harvard \& Smithsonian.}
Karin I. \"Oberg,$^{2}$
Guillermo M. Mu\~noz Caro,$^{1}$
H\'ector Carrascosa,$^{1}$
\newauthor Asunci\'on Fuente$^{1}$, 
Mahesh Rajappan$^{2}$\\
$^{1}$Centro de Astrobiolog\'ia (CSIC-INTA) 
Carretera de Ajalvir, km. 4, Torrej\'on de Ardoz, E-28850, Madrid, Spain\\
$^{2}$Center for Astrophysics $|$ Harvard \& Smithsonian, 60 Garden St., Cambridge, MA 02138, USA\\
}

\date{Accepted XXX. Received YYY; in original form ZZZ}

\pubyear{2024}

\begin{document}
\label{firstpage}
\pagerange{\pageref{firstpage}--\pageref{lastpage}}
\maketitle

\begin{abstract}
Understanding the formation of carbonyl sulfide (OCS) in interstellar ices is key to constrain the sulfur chemistry in the interstellar medium (ISM), since it is the only ice S-bearing molecule securely detected thus far. 
Two general pathways for OCS formation have been proposed: sulfurization of CO (CO+S) and oxidation of CS (CS+O), but their relative contribution in interstellar ices remains unconstrained.
We have evaluated the contribution of both pathways to OCS formation upon energetic processing in isotopically-labeled CO$_2$:CS$_2$ and CO:CS$_2$ ice samples at 7$-$50 K. 
Our results indicated that formation of OCS through the CS+O pathway was more favorable than through the CO+S pathway, as previously suggested by theoretical calculations. 
%
In addition, its relative contribution increased at higher temperatures. 
Therefore, this pathway could play a role in the ice formation of OCS, especially in warm regions where CO is expected to be preferentially in the gas phase. 
%
%
At the same time, we have explored the chemistry of CS$_2$-bearing, CO$_2$-, CO-, and also H$_2$O-rich ices, that could be relevant to the sulfur interstellar chemistry.
%
%
We observed formation of a variety of S-bearing products in addition to OCS, including SO$_2$, C$_3$S$_2$, and S$_2$. 
However, a significant fraction of sulfur 
was not detected at the end of the experiments, and could be locked in long, undetectable sulfur allotropes, one of the potential carriers of the missing sulfur in the dense ISM. 
 
\end{abstract}

\begin{keywords}
astrochemistry;
solid state: volatile;
ISM: clouds;
ISM: molecules
\end{keywords}



\section{Introduction}\label{sec:intro}
Sulfur (S) is the tenth most abundant element in the universe \citep[S/H $\sim$ 1.5 $\times$ 10$^{-5}$,][]{asplund09}, and it is present in a variety of biomolecules on Earth (amino acids, nucleic acids, vitamins). 
Despite its relevance, the fate of sulfur in the interstellar medium (ISM) once it is incorporated into the interior of dense clouds is uncertain \citep[see, e.g.,][and references therein]{laas19}. 
%
Most of the sulfur in the dense ISM is expected to be locked in the solid phase, either on ice mantles or in (semi-)refractory form \citep{millar90,ruffle99,vidal17,drozdovskaia18,kama19,fuente19,pablo20,fuente23}. 
However, detections of solid sulfur in these regions have been sparse. 
In interstellar ices, only carbonyl sulfide (OCS) has been securely detected 
\citep[see, e.g.,][]{boogert22,mcclure23}, while the presence of sulfur dioxide (SO$_2$) has also been suggested \citep[most recently in][]{rocha23}. 
Unfortunately, their reported abundances only account for $<$5\% of the cosmic sulfur abundance. 

Even though they do not represent the main sulfur carrier in the dense ISM, understanding how this reservoir of sulfur in interstellar ices is built would be of vital importance to constrain the sulfur chemistry in the ISM. 
In the case of OCS, direct accretion from the gas phase is probably not significant, based on its low gas-phase abundance with respect to CO compared to the OCS/CO ratio measured in ices \citep[see, e.g.,][]{boogert22}. Therefore, \textit{in situ} formation on the grain surface or the ice bulk is usually invoked. 
Two general pathways for OCS formation have been presented in the literature: sulfurization of CO (CO+S) and oxidation of CS (CS+O). 
\citet{adriaens10} theoretically studied the atom-addition reactions CO + S and CS + O, as well as the neutral-radical reactions CO + HS and CS + OH, both in the gas phase and on coronene (as an analog to carbonaceous dust grains). 
The results showed that for both, the atom-addition and the neutral-radical reactions, oxidation of CS was more energetically favorable than sulfurization of CO. 
In any case, the authors indicated that the dominant OCS formation route in the ISM would depend not only on the efficiency and rate of the reactions, but also on the availability of the reactants. 

\smallskip
Formation of OCS in interstellar ices has been studied in the laboratory under astrophysically relevant conditions. 
The sulfurization of CO was first explored in \citet{ferrante08} and \citet{garozzo10}. 
These authors performed 800 and 200 keV (respectively) proton irradiation experiments at 10$-$20 K of ice samples containing CO and either SO$_2$ or hydrogen sulfide (H$_2$S) as the S-source. 
H$_2$S has not yet been detected in interstellar ices, but it is the most abundant S-bearing molecule in comets \citep[see, e.g.,][]{calmonte16}. 
OCS formation was observed in both cases. 
%
\citet{ferrante08} also performed experiments with CO$_2$ as the CO source and observed a decrease in the OCS formation efficiency. 
This suggested that dissociation of the CO$_2$ molecules (in addition to dissociation of the S-carrier) was required prior to the formation of OCS through the atom-addition reaction 
\begin{equation}
    \centering
    \mathrm{CO + S \rightarrow OCS}.\label{eq:co+s}
\end{equation}
%
%
%
%
%

Irradiation of CO:H$_2$S and CO$_2$:H$_2$S ice samples at 14 K was carried out in \citet{asper15} using vacuum and extreme ultraviolet (VUV and EUV) photons. 
The results were similar to those reported in \citet{ferrante08}. 
%
%
Experiments with different initial ice compositions revealed that formation of OCS competed with formation of H$_2$S$_2$ (the latter being favored in samples with higher H$_2$S abundances). 
Therefore, \citet{asper15} suggested that, when H$_2$S is the S-source, formation of OCS may proceed through both, the atom-addition (Eq. \ref{eq:co+s}) and the neutral-radical reaction:
\begin{equation}
    \centering
\mathrm{CO + HS \rightarrow OCS + H}.\label{eq:co+hs}
\end{equation}
The CO + HS reaction has recently been revisited in \citet{elakel22} during co-deposition of CO, H$_2$S, and H atoms on a substrate at 10 K and 22 K.
According to these works, the neutral-radical reaction takes place in two steps: 
\begin{equation}
    \centering
\mathrm{CO + HS \rightarrow HSCO}\label{eq:co+hs_2}
\end{equation}
\begin{equation}
    \centering
\mathrm{HSCO + H \rightarrow OCS + H_2.}\label{eq:hsco+h}
\end{equation}
%

On the other hand, the oxidation of CS was explored in \citet{ward12} and \citet{maity13} using carbon disulfide (CS$_2$) as the CS source in both cases. 
Even though CS$_2$ is readily formed in the laboratory upon irradiation of CO:H$_2$S ice mixtures \citep[see][]{ferrante08,garozzo10,asper15}, this molecule has not yet been detected in interstellar ices, perhaps due to its subsequent conversion to other S-bearing species \citep{ward12,maity13}. In the Solar System, CS$_2$ has been detected in comets \citep{calmonte16} and in planetary bodies such as the Jovian system \citep{noll95}. Therefore, studying the CS$_2$ ice chemistry could also be relevant to constrain the sulfur interstellar chemistry.  
\citet{ward12} studied the thermal reaction between simultaneously deposited CS$_2$ molecules and O atoms on a surface at 15$-$70 K. 
The formation of OCS below 25 K mainly took place through a two-step mechanism:

\begin{equation}
\centering
\mathrm{CS_2 + O \rightarrow CS + SO}
\end{equation} 

\begin{equation}
\centering
\mathrm{CS + O \rightarrow OCS.}\label{eq:cs+o}
\end{equation}

%
%
%
 
\citet{maity13} studied the formation of OCS upon 5 keV electron irradiation of CS$_2$:O$_2$ ice mixtures at 12 K. The authors proposed a comprehensive reaction scheme that included the formation of additional S-bearing (SO$_2$, SO$_3$, C$_3$S$_2$) and non-S-bearing (CO, CO$_2$, O$_3$) molecules detected by IR spectroscopy.
In particular, formation of OCS took place through the atom-addition reaction of ground (singlet) state CS with suprathermal O atoms in the ground (triplet) state:

\begin{equation}
\centering
\mathrm{CS_2(X^1\Sigma^+_g)  \rightarrow CS(X^1\Sigma^+) + S(^3P)}\label{eq:cs2}
\end{equation} 

\begin{equation}
\centering
\mathrm{O_2(X^3\Sigma^-_g)  \rightarrow O(^3P) + O(^3P)}
\end{equation} 

\begin{equation}
\centering
\mathrm{CS_2(X^1\Sigma^+_g) + O(^3P)  \rightarrow CS(X^1\Sigma^+) + SO(X^3\Sigma^-) }
\end{equation} 

\begin{equation}
\centering
\mathrm{CS(X^1\Sigma^+) + O(^3P)  \rightarrow OCS(X^1\Sigma^+).}\label{eq:cs+o2}
\end{equation}


A similar set of reactions would be expected if a different O-bearing molecule such as H$_2$O or CO$_2$ acted as a source of O atoms \citep{maity13}. 
We note that reaction \ref{eq:cs+o2} (and also reactions \ref{eq:co+s} and \ref{eq:cs+o} when reactants are in the ground electronic state) are, in principle, spin-forbidden. \citet{adriaens10} indicated that such reactions would initially form OCS in the excited triplet state, which would be stabilized on the coronene surface and relaxed to the ground singlet state. A similar behavior would be expected in the ice. 
Likewise, \citet{okabe78} suggested that reaction \ref{eq:cs2} may take place in violation of the spin conservation rules because of the presence of heavy S atoms. 

Previous laboratory experiments addressing the ice formation of OCS can thus be divided in two groups: 
1) those studying the sulfurization of CO in ice samples containing CO or CO$_2$ and a S-source (either SO$_2$ or H$_2$S), and 
2) those studying the oxidation of CS in ice samples containing CS$_2$ and a O-source (either O atoms or irradiated O$_2$ 
molecules). 
\citet{sivaraman16} reported the formation of OCS molecules in 2 keV electron irradiated CO$_2$:CS$_2$ ice samples at 85 K, but could not pinpoint whether they formed through the CO+S and/or CS+O formation pathways. 
In this work, we have simultaneously studied both OCS formation pathways upon 2 keV electron irradiation of ice samples containing isotopically-labeled CO or CO$_2$ (as a source of CO and O atoms) and CS$_2$ (as a source of CS and S atoms) at 7$-$50 K.  
These experiments have allowed us to 
i) compare the relative contribution of the CO+S and CS+O pathways in samples where both are accessible, and 
ii) constrain the effect of ice temperature on said pathways. 
Additional irradiation experiments of H$_2$O:CS$_2$ and CO:H$_2$S ice samples were also performed to explore both pathways in different ice environments. 
With these experiments we have also been able to evaluate the sulfur chemistry of ices containing CS$_2$ in realistic environments (H$_2$O, CO, and CO$_2$ ice matrices), which could be relevant to the sulfur interstellar chemistry.
The experimental setup used for the laboratory experiments is described in Sect. \ref{sec:exp}, and the results are presented in Sect. \ref{sec:results}. 
Sections \ref{sec:disc_ocs} and \ref{sec:disc_ocsT} discuss the relative contribution of the two OCS formation pathways and their dependence with the ice temperature, while the chemistry of CS$_2$-bearing ices is discussed in Sect. \ref{sec:disc_chem}. Finally, Sect. \ref{sec:disc_astimp} summarizes the astrophysical implications of this work, and Sect. \ref{sec:conc} lists the main conclusions.

\section{Methods}\label{sec:exp}
\begin{table*}
    \centering
    \begin{tabular}{cccccc}
       Exp. & Ice comp. & Comp. ratio & Ice thickness (ML) & Irrad. T (K) & Irrad. Fluence ($\times$10$^{18}$ eV cm$^{-2}$) \\
       \hline
       1 & $^{13}$C$^{18}$O$_2$:CS$_2$ & 88:12 & 205 & 7 &8.5\\ 
       2 & $^{13}$C$^{18}$O$_2$:CS$_2$ & 85:15 & 205 & 8 &8.8\\ 
       3 & $^{13}$C$^{18}$O$_2$:CS$_2$ & 84:16 & 195 & 25 &8.6\\
       4 & $^{13}$C$^{18}$O$_2$:CS$_2$ & 84:16 & 205 & 50 &8.9\\
       5 & $^{13}$C$^{18}$O$_2$:CS$_2$ & 93:7 & 530 & 10 &4.5$^*$\\ 
       6 & $^{13}$C$^{18}$O$_2$:CS$_2$ & 93:7 & 490 & 11 &4.3$^*$\\
       \hline
       7 & $^{13}$C$^{18}$O:CS$_2$ & 88:12 & 205 & 7 &8.6\\
       8 & $^{13}$C$^{18}$O:CS$_2$ & 89:11 & 210 & 15 &7.4\\
       9 & $^{13}$C$^{18}$O:CS$_2$ & 89:11 & 220 & 25 &8.0\\
       \hline
       10 & H$_2$$^{18}$O:CS$_2$ & 74:26 & 340 & 8 & 8.4\\
       11 & H$_2$$^{18}$O:CS$_2$ & 74:26 & 340 & 50 & 8.1\\
       \hline
       12 & CO:H$_2$S & 82:18 & 285 & 6 & 7.7\\
       13 & CO:H$_2$S & 80:20 & 270 & 25 & 8.2\\
       \hline
    \end{tabular}
    \caption{Summary of the performed irradiation experiments. 
    The ice composition ratios and thicknesses were measured at the deposition temperature (6$-$8 K). 
    A 30\% and 40\% uncertainty (respectively) should be considered for these values (see the text). 
    Note that 1 ML = 10$^{15}$ molecules cm$^{-2}$.
    $^*$Photon irradiation instead of electron irradiation. 
    }
    \label{tab:exp}
\end{table*}

Table \ref{tab:exp} lists the irradiation experiments carried out for this paper. 
The resulting ice chemistry products were detected by a combination of IR spectroscopy of the ice samples (Sect. \ref{sec:exp_IR}) and quadrupole mass spectrometry of the desorbing molecules during temperature programmed desorption (TPD) of the irradiated ices (Sect. \ref{sec:exp_TPD}).
In order to distinguish between the OCS molecules formed through the CO+S and the CS+O pathways in Experiments 1$-$9, we isotopically labeled the C in the initial CO and CO$_2$ molecules. 
Therefore, the OCS molecules formed through the CO+S pathway contained $^{13}$C, while those formed through the CS+O pathway did not. 
In addition, we isotopically labeled the O in the CO, CO$_2$, and H$_2$O molecules in Experiments 1$-$11 in order to shift the position of the IR bands and avoid overlapping with other features.   
As a result, the oxidation of CS led to the formation of $^{18}$OCS (with an IR feature at $\sim$2010 cm$^{-1}$ and a molecular mass of 62 amu), while the sulfurization of CO led to the formation of $^{18}$O$^{13}$CS (with an IR feature at $\sim$1955 cm$^{-1}$ and a molecular mass of 63 amu). 

Experiments 1$-$4 and 7$-$13 were carried out using the SPACE TIGER experimental setup, consisting of a ultra-high-vacuum (UHV) chamber with a base pressure of $\sim$2 $\times$10$^{-10}$ Torr that is relevant to dense cloud interior conditions. 
The ice samples in Experiments 5 and 6 were irradiated with VUV photons instead of 2 keV electrons to explore the influence of the energy source in the formation of OCS molecules. These experiments were performed in the SPACE CAT setup. 
These setups are described in detail in \citet{pavlo22} and \citet{lauck15}, respectively. In this Section we present those features specific to the reported experiments. 

\subsection{Ice sample preparation}
The ice samples in Experiments 1$-$4 and 7$-$13 were deposited on a copper substrate at 6$-$8 K located at the center of the UHV chamber, and warmed up to the corresponding irradiation temperature when needed (fifth column of Table \ref{tab:exp}). 
The ices in Experiments 5 and 6 were deposited on a CsI substrate at 10$-$11 K and irradiated at the same temperature. 
In both cases, the ice samples were grown by exposing the substrate to a gas mixture with the desired composition, introduced in the chamber from an independently pumped gas line assembly. The gas mixtures were composed by a combination of  $^{13}$C$^{18}$O$_2$ (gas, 95\%, Sigma-Aldrich), $^{13}$C$^{18}$O (gas, 99\%, Sigma-Aldrich), CO (gas, 99.95\%, Sigma-Aldrich), CS$_2$ (anhydrous, $\ge$99\%, Sigma-Aldrich), H$_2$$^{18}$O (liquid, 99\%, Sigma-Aldrich), and H$_2$S (gas, $\ge$99.5\%, Sigma-Aldrich). 
CS$_2$ and H$_2$$^{18}$O were used after applying three freeze-thaw-pump cycles. 
The composition and thickness of the ice samples were determined by IR spectroscopy (Sect. \ref{sec:exp_IR}), and are indicated in the third and fourth columns of Table \ref{tab:exp}, respectively. 

\subsection{2 keV electron irradiation of the ice samples}\label{sec:exp_e}
The ice samples in Experiments 1$-$4 and 7$-$15 were electron irradiated using an ELG-2/EGPS-1022 electron source system provided by Kimball Physics. The energy of the irradiated electrons was 2 keV. 
In the dense ISM, cosmic rays (consisting predominantly of protons) interacting with interstellar ices lose most of their kinetic energy via transfer to the electronic system of the target molecules. This produces the so-called $\delta$-electrons within the ice mantles, with energies of up to a few keV \citep{jones11}. The corresponding electronic linear energy transfer is on the order of a few keV $\mu$m$^{-1}$ \citep{hovington97}, similar to that corresponding to the irradiation of ice samples in the laboratory using keV electrons \citep{bennett04}. 
In the reported experiments the electron beam current varied between 90 and 110 nA, with an average irradiation time of $\sim$110 minutes, leading to a total incident energy fluence of 7.3$-$8.9 $\times$ 10$^{18}$ eV cm$^{-2}$ (sixth column of Table \ref{tab:exp}). This would correspond to the energy deposited by the cosmic rays into the interstellar ice mantles in the interior of dense clouds during $\sim$5 $\times$ 10$^6$ years \citep[see][and references therein]{jones11}. 

The penetration depth of the 2 keV electrons (defined as the ice thickness where 75\% of the electron energy was lost) depended on the composition of the ice samples, and was calculated with the CASINO v2.42 code \citep{drouin07}.
This value was 
$\sim$180 ML (1 ML = 10$^{15}$ molecules cm$^{-2}$) for the $^{13}$C$^{18}$O$_2$:CS$_2$ ice samples in Experiments 1$-$4, 
$\sim$170 ML for the $^{13}$C$^{18}$O:CS$_2$ ice samples in Experiments 7$-$9,  
$\sim$260 ML for the H$_2^{18}$O:CS$_2$ ice samples in Experiments 10$-$11, 
and $\sim$195 ML for the CO:H$_2$S ice samples in Experiments 12$-$13. 

\subsection{VUV photon irradiation of the ice samples}
The ice samples in Experiments 5 and 6 were irradiated with VUV photons using a Hamamatsu H2D2 L11798 deuterium lamp. 
The lamp emission spectrum is presented in \citet{bergner17} and \citet{martin20}, and features a strong emission band at $\sim$160 nm and a weaker emission around Ly$\alpha$ wavelengths. This resembles the secondary UV field produced in the interior of dense clouds from the interaction of cosmic rays with gas-phase H$_2$ molecules \citep{pestellini92,shen04}. The mean energy of the irradiated VUV photons was 7.8 eV. 
The VUV flux at the sample position in Experiments 5 and 6 was measured with a NIST calibrated AXUV-100G photodiode placed in front of the substrate (5.3 $\times$ 10$^{13}$ and 5.1 $\times$ 10$^{13}$ photons cm$^{-2}$ s$^{-1}$, respectively). The irradiation time was 180 minutes in both experiments, leading to a total fluence of 5.7 $\times$ 10$^{17}$ and 5.5 $\times$ 10$^{17}$ photons cm$^{-2}$, respectively. This corresponded to an incident energy fluence of 4.5 $\times$ 10$^{18}$ and 4.3 $\times$ 10$^{18}$ eV cm$^{-2}$ (sixth column of Table \ref{tab:exp}). For comparison, ice mantles in the interior of dense clouds are exposed to a fluence of $\sim$6 $\times$ 10$^{17}$ photons cm$^{-2}$ during 2 $\times$ 10$^6$ years, assuming an interstellar secondary UV flux of $\sim$10$^4$ photons cm$^{-2}$ s$^{-1}$ \citep{pestellini92,shen04}. 

The average VUV-absorption cross section of CO$_2$ ice in the 120$-$160 nm range is 6.7 $\times$ 10$^{-19}$ cm$^{-2}$ \citep{gustavo14b}, meaning that 4450 ML of CO$_2$ ice are needed to absorb 95\% of the incident photons in that range. Unfortunately, the VUV-absorption cross section of CS$_2$ ice has not yet been reported. Therefore, it was not possible to estimate the absorbed energy in Experiments 5 and 6.

\subsection{IR ice spectroscopy}\label{sec:exp_IR}

\begin{table*}
    \centering
    \begin{tabular}{cccc}
     Molecule&Wavenumber&Band strength&Reference\\
      & (cm$^{-1}$) & (cm molecule$^{-1}$) & \\
     \hline
     H$_2$O & 760 & 3.1 $\times$10$^{-17}$ & \citet{gerakines95}\\
     CO$_2$ & 2343 & 7.6 $\times$10$^{-17}$ & \citet{gerakines95}\\
     CO$_2$ & 660 & 1.1 $\times$10$^{-17}$ & \citet{gerakines95}\\
     CO & 2139 & 1.1 $\times$10$^{-17}$ & \citet{gerakines95}\\
     C$_3$O$_2$ & 2242 & 1.3 $\times$10$^{-16}$ & \citet{gerakines01}\\
     CS$_2$ & 1501 & 1.1 $\times$10$^{-16}$ & \citet{angele24}\\
     H$_2$S & 2547 & 1.7 $\times$10$^{-17}$ & \citet{yarnall22}\\
     OCS & 2031 & 1.2  $\times$10$^{-16}$ & \citet{yarnall22}\\
     SO$_2$ & 1323 & 4.2 $\times$10$^{-17}$ & \citet{yarnall22}\\ 
     SO$_3$ & 1385 & 3.0 $\times$10$^{-17}$ & \citet{majkowski78}\\ 
    \end{tabular}
    \caption{Band strengths of selected features in pure ice IR spectra collected in transmittance. 
    The same band strengths were assumed for the different isotopologs of every species as a first approximation, which could introduce an uncertainty of up to 20\% \citep{gerakines95}. We used the 660 cm$^{-1}$ IR feature instead of the main feature at 2343 cm$^{-1}$ to estimate the $^{13}$C$^{18}$O$_2$ ice column density in Experiments 1$-$6, because it appeared in a cleaner region of the spectrum. 
    }
    \label{tab:ir}
\end{table*}

The ice samples in Experiments 1$-$4 and 7$-$13 were monitored through reflection-absorption IR spectroscopy (RAIRS), while those in Experiments 5 and 6 were monitored through IR spectroscopy in transmittance. In both cases we used a Bruker 70v Fourier transform IR (FTIR) spectrometer with a liquid-nitrogen-cooled MCT detector. 
The spectra were averaged over 256 interferograms and collected with a resolution of 1 cm$^{-1}$ in the 5000$-$600 cm$^{-1}$ range. 

The integrated absorbances of the detected IR features were used to estimate the ice column densities ($N$) in molecules cm$^{-2}$ of the corresponding species using the equation: 

\begin{equation}
N=\frac{1}{A}\int_{band}{\tau_{\nu} \ d\nu},
\label{eqn}
\end{equation}

\noindent where $\tau_{\nu}$ is the optical depth of the absorption band (2.3 times the absorbance), and $A$ is the band strength of the IR feature in cm molecule$^{-1}$.
The IR features were numerically integrated using the \texttt{integrate.simps} function in the SciPy library. 
When neighboring bands overlapped, the corresponding region of the IR spectrum was fitted with multiple Gaussians using the \texttt{curve$\_$fit} function. 
The integrated absorbances were then calculated as the area of the Gaussians. 
Table \ref{tab:ir} lists the reported band strengths in transmittance of selected IR features in pure ices of the main isotopologs. 
%
In this work we assumed the same band strengths for the different isotopologs of the same species as a first approximation, 
which could introduce an uncertainty of up to 20\% \citep{gerakines95}. 
In addition, the band strengths in reflection-absorption IR spectra differ from those in transmittance spectroscopy, and are setup specific. 
In the case of SPACE TIGER, \citet{martin24} reported a ratio of $\sim$2.3 for the 4253 cm$^{-1}$ CO IR band strength in the reflection-absorption and transmittance modes, with a 35\% uncertainty. 
This ratio has been adopted for all IR features in the present work. 
As a result, we estimated an absolute uncertainty of 40\% in the ice column densities calculated from the IR integrated absorbances in Experiments 1$-$4 and 7$-$13. 
We note that only a 30\% uncertainty was considered when calculating column density ratios (due to the 20\% uncertainty of the individual column densities), because the $\sim$2.3 ratio (carrying an additional 35\% uncertainty, as explained above) was not used in that case. 
The initial ice thicknesses and composition ratios are listed in the third and fourth columns of Table \ref{tab:exp}. 

\subsection{Temperature Programmed Desorption (TPD) of the irradiated ice samples}\label{sec:exp_TPD}
The irradiated ice samples were warmed from the irradiation temperature (fifth column in Table \ref{tab:exp}) up to 250 K at a controlled heating rate of 2 K min$^{-1}$. 
The molecules desorbing from the ice mantles to the gas phase were detected with a Pfeiffer QMG 220M1 QMS (mass-to-charge range of 1$-$100 amu and 0.5 amu resolution). 
To this purpose, we monitored the signal corresponding to mass-to-charge ratios ($m/z$) equal to the molecular mass (and/or to relevant molecular fragments) of the initial ice components and the expected ice chemistry products. 
The evolution of the QMS signal of desorbing species as a function of temperature is known as a TPD curve. 

For those species with no detectable IR features, calibration of the QMS allowed a rough estimation of the desorbing column density  ($N(X)$) from the integrated QMS signal of the corresponding TPD curve ($A(m/z)$) using Eq. \ref{eqn_qms} \citep{martin15} : 

\begin{equation} 
N(X) = \frac{A(m/z)}{k_{CO}} \cdot \frac{\sigma^+(CO)}{\sigma^+(X)} 
\cdot \frac{F_F(28)}{F_F(m)} \cdot \frac{S(28)}{S(m/z)},  \label{eqn_qms}
\end{equation}

where 
$k_{CO}$ was the proportionality constant for CO molecules, 
$\sigma^+(X)$ was the electron-impact ionization cross-section for species $X$, 
$F_F(m)$ was the fraction of molecules $X$ leading to a fragment of mass $m$ in the QMS,  
and $S(m/z)$ was the sensitivity of the QMS to the mass fragment $m/z$. 
The $k_{CO}$ and $S(m/z)$ parameters were derived from dedicated calibration experiments described in detail in \citet{martin24}. 
New calibration experiments were performed to update the $k_{CO}$ value (1.15 $\times$ 10$^{-11}$ A K ML$^{-1}$), 
and the equations to calculate $S(m/z)$: 

\begin{equation}
    k^*_{QMS} \cdot S(m/z) = 2.31 \times 10^{15} \cdot e^{\frac{-m/z}{12.38}} (m/z < 40)
\end{equation}

\begin{equation}
    k^*_{QMS} \cdot S(m/z) = 2.37 \times 10^{14} \cdot e^{\frac{-m/z}{41.72}} (m/z > 40) 
\end{equation} 

(where $k^*_{QMS}$ was a different proportionality constant that canceled out when calculating the $S(28)/S(m/z)$ ratio). 
A 50\% uncertainty was estimated in \citet{martin24} for the column densities calculated with Eq. \ref{eqn_qms}. 

\section{Results}\label{sec:results}

The 2 keV electron irradiation of $^{13}$C$^{18}$O$_2$:CS$_2$ and $^{13}$C$^{18}$O:CS$_2$ ice samples at 7 K  (Experiments 1, 2, and 7) are presented in Sections \ref{sec:co2_cs2_7K} and \ref{sec:co_cs2_7K}, respectively. 
These two sections first introduce an overview of the ice chemistry (with special attention to the S-bearing molecules), and then focus on the relative contribution of the CO+S and CS+O pathways to the formation of OCS. 
Section \ref{sec:co2_cs2_vuv} presents the VUV photon irradiation of $^{13}$C$^{18}$O$_2$:CS$_2$ samples (Experiments 5 and 6), aimed at evaluating the effect of the energy source on the formation of OCS. 
Sections \ref{sec:co2cs2_T} and \ref{sec:cocs2_T} assess the effect of the ice temperature on the relative contributions of the two OCS formation pathways in $^{13}$C$^{18}$O$_2$:CS$_2$ and $^{13}$C$^{18}$O:CS$_2$ samples, respectively (Experiments 3, 4, 8, 9). 
Finally, Sections \ref{sec:h2o_cs2} and \ref{sec:co_h2s} explore whether the same trends observed in previous sections apply to different ice compositions depicting other plausible astrophysical environments (Experiments 10$-$13). 

\subsection{2 keV electron irradiation of a $^{13}$C$^{18}$O$_2$:CS$_2$ ice sample at 7 K}\label{sec:co2_cs2_7K}

\begin{figure}
    \centering
    \includegraphics[width=1\linewidth]{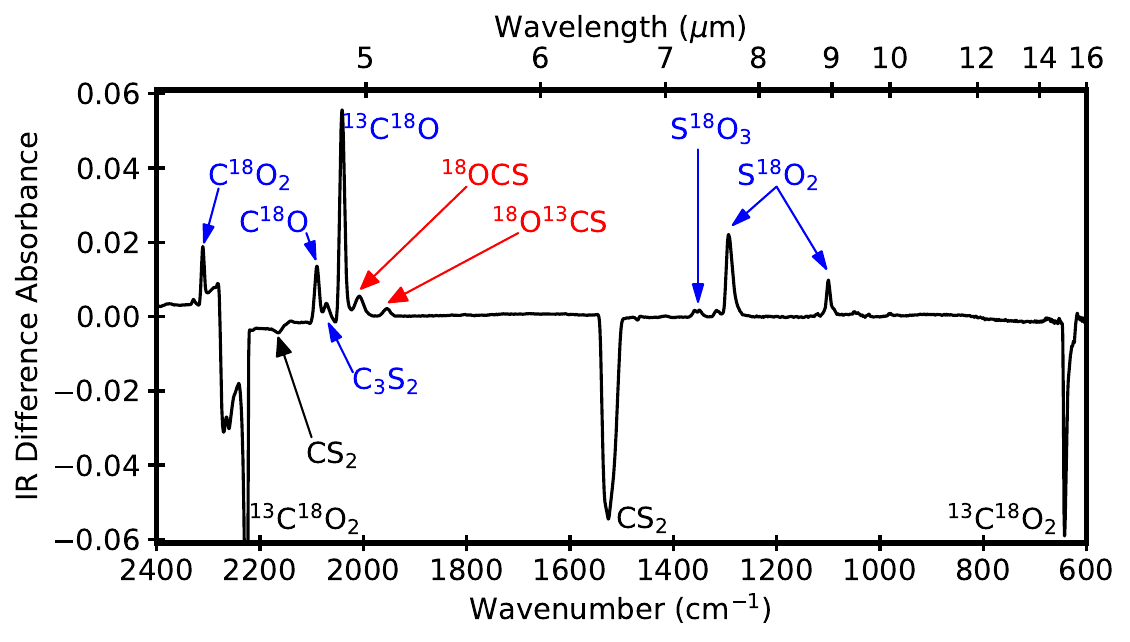}
    \caption{IR difference spectrum obtained upon 2 keV electron irradiation of a $^{13}$C$^{18}$O$_2$:CS$_2$ ice sample at 7 K in Exp. 1. IR band assignments are indicated for the initial ice components (black), and ice chemistry products (blue), including the OCS isotopologs (red).}
    \label{fig:co2cs2_IR}
\end{figure}

\begin{figure*}
    \centering
    \includegraphics[width=0.6\linewidth]{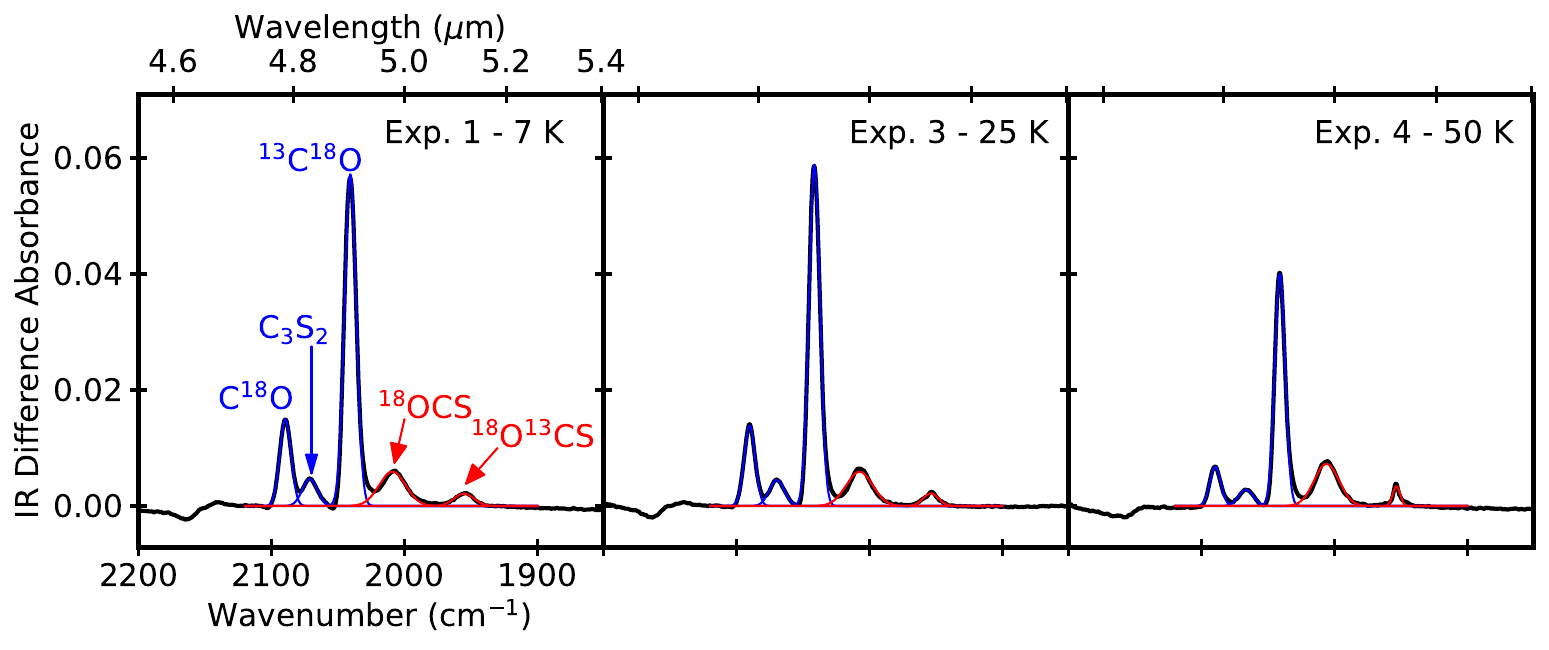}
    \includegraphics[width=0.61\linewidth]{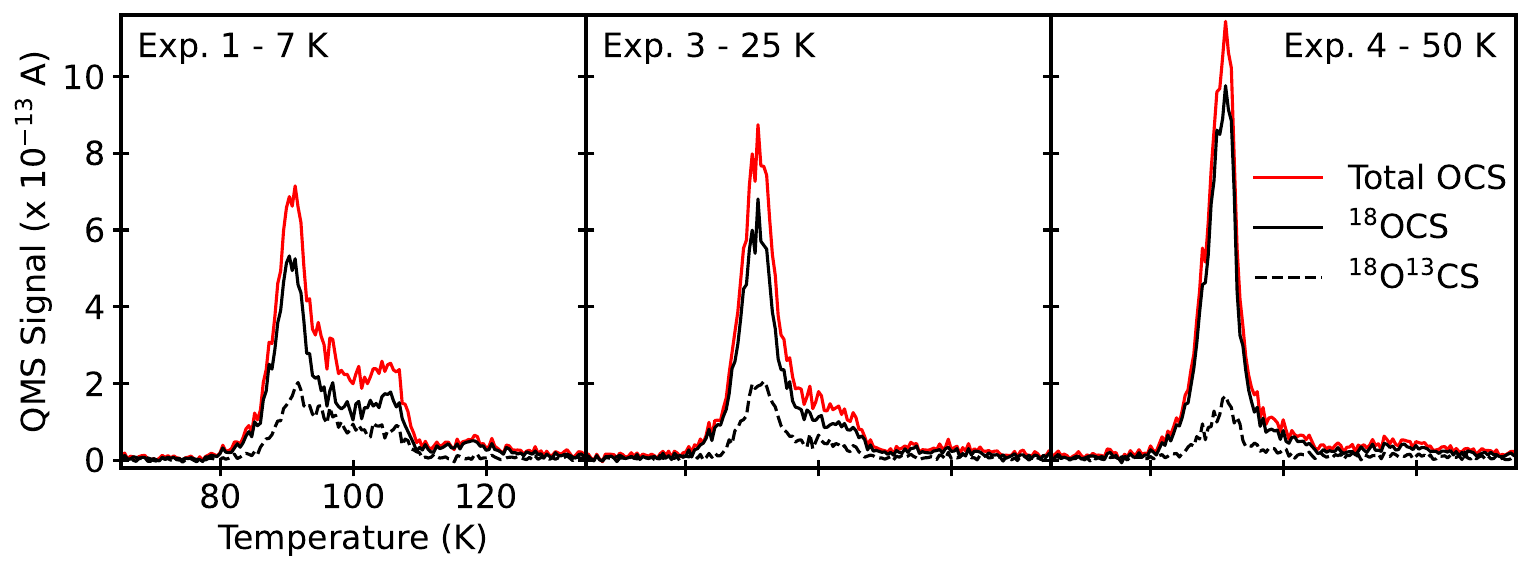}
    \caption{\textit{Top}: Gaussian fitting of the features corresponding to C$^{18}$O, C$_3$S$_2$, $^{13}$C$^{18}$O (blue), $^{18}$OCS and $^{18}$O$^{13}$CS (red) in the IR difference spectra of the $^{13}$C$^{18}$O$_2$:CS$_2$ ice samples irradiated at 7 K (left panel), 25 K (middle panel), and 50 K (right panel).
    \textit{Bottom}: TPD curves corresponding to $^{18}$OCS (solid black) and $^{18}$O$^{13}$CS (dashed black), along with the sum of both signals (solid red), measured during thermal desorption of the irradited ice samples in the same experiments. 
    }
    \label{fig:co2cs2_tpd_ocs}
    \label{fig:co2cs2_gauss}
\end{figure*}

\begin{figure}
    \centering
    \includegraphics[width=0.55\linewidth]{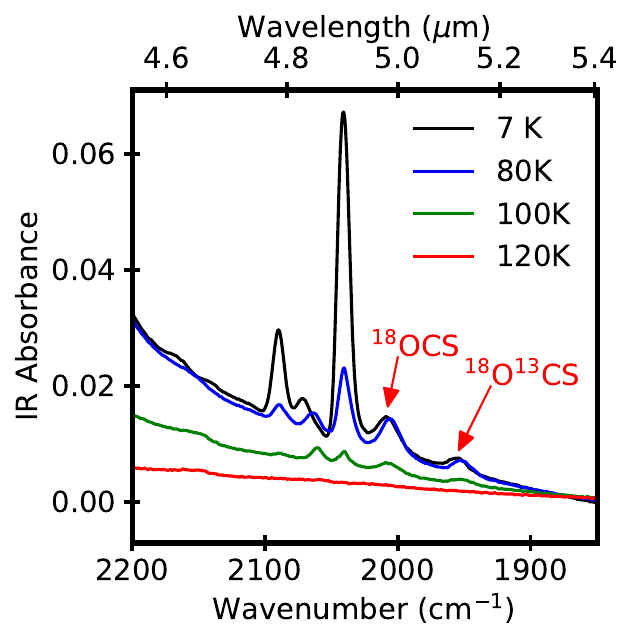}
    \caption{Evolution of the IR spectrum in the 2200$-$1850 cm$^{-1}$ range during the TPD of the irradiated $^{13}$C$^{18}$O$_2$:CS$_2$ ice sample in Exp. 1.}
    \label{fig:co2cs2_7K_IR_evo}
\end{figure}

Fig. \ref{fig:co2cs2_IR} shows the IR difference spectrum in the 2400$-$600 cm$^{-1}$ range obtained upon 2 keV electron irradiation at 7 K of the $^{13}$C$^{18}$O$_2$:CS$_2$ ice sample in Exp. 1. 
Results were comparable in Exp. 2, with a similar initial ice composition and irradiated energy fluence, 
(see top panel of Fig. \ref{fig:co2_cs2_IR_allT} in Appendix \ref{app:co2_cs2})
Energetic processing of the ice led to dissociation of $^{13}$C$^{18}$O$_2$ and CS$_2$ molecules, resulting in negative features in Fig. \ref{fig:co2cs2_IR}, while formation of new products led to positive features. 
According to their corresponding IR absorbances, $\sim$27\% of the initial $^{13}$C$^{18}$O$_2$ and $\sim$60\% of the initial CS$_2$ molecules were depleted after irradiation. 
%
The main irradiation product was $^{13}$C$^{18}$O (detected at $\sim$2040 cm$^{-1}$), most likely formed through dissociation of the $^{13}$C$^{18}$O$_2$ molecules.  
%
C$^{18}$O molecules were also detected at $\sim$2090 cm$^{-1}$, and could be formed through dissociation of produced $^{18}$OCS molecules \citep{maity13} or through C + $^{18}$O atom-addition reactions.
Further oxidation of the C$^{18}$O molecules led to the formation of C$^{18}$O$_2$, detected at $\sim$2310 cm$^{-1}$. 
We note that formation of $^{13}$C$^{18}$O$_3$ and $^{18}$O$_3$, usually detected upon energetic processing of CO$_2$-bearing ices \citep[see, e.g.,][]{ferrante08,asper15}, was not significant in Experiments 1 and 2. 
In particular, even though a very weak IR feature was tentatively observed at 983 cm$^{-1}$, its assignment to $^{18}$O$_3$ \citep{maity13} could not be confirmed because thermal desorption of this species was not detected during the subsequent TPD. 
Therefore, this feature was not marked in Figures \ref{fig:co2cs2_IR} and \ref{fig:co2_cs2_IR_allT}. 
The $^{13}$C$^{18}$O$_3$ IR feature, on the other hand, could be overlapped with the $^{18}$O$^{13}$CS IR band (see Sect. \ref{sec:co2_cs2_vuv}). However, the $^{18}$OCS/$^{18}$O$^{13}$CS ratio calculated from the IR absorbances was similar to that obtained from the TPD curves (see below). Since the $m/z$ 63 TPD curve corresponding to $^{18}$O$^{13}$CS could not present any contribution from $^{13}$C$^{18}$O$_3$ (with a molecular mass of 67 amu), we assumed that the contribution of $^{13}$C$^{18}$O$_3$ to the $^{18}$O$^{13}$CS IR feature was negligible.
This could indicate that $^{18}$O atoms resulting from the dissociation of $^{13}$C$^{18}$O$_2$ molecules preferentially reacted with CS$_2$ and its dissociation products.

On the other hand, the main detected S-bearing product was S$^{18}$O$_2$, with two IR features detected at 1290 and 1100 cm$^{-1}$ \citep{maity13}.  
Thermal desorption of this molecule at T $\sim$ 110 K was detected during the TPD of the irradiated samples, and is shown in Appendix \ref{app:co2_cs2} (bottom left panel of Fig. \ref{fig:co2cs2_tpd_so2}). 
The estimated S$^{18}$O$_2$ ice column density after irradiation was $\sim$8 ML.   
This accounted for 
$\sim$16\% of the initial sulfur in the ice samples, 
or $\sim$26\% of the sulfur consumed during irradiation. 
%
Oxidation of S$^{18}$O$_2$ molecules led to the formation of $\sim$1 ML of sulfur trioxide (S$^{18}$O$_3$),  
detected at 1350 and 1357 cm$^{-1}$ \citep{maity13}. 
S$^{18}$O$_3$ accounted for $\sim$2\% of the initial sulfur, 
or $\sim$4\% of the consumed sulfur. 


Formation of OCS molecules took place through both, the CS+O and the CO+S pathways, leading to the detection of two IR features at 2009 and 1955 cm$^{-1}$ (corresponding to $^{18}$OCS and $^{18}$O$^{13}$CS, respectively). 
The $^{18}$OCS IR feature was detected at 2002 cm$^{-1}$ in \citet{maity13}, while the position of the $^{18}$O$^{13}$CS feature had not been previously reported in the literature. 
The assignment of both features was confirmed during the TPD of the irradiated ice sample. 
The bottom left panel of Fig. \ref{fig:co2cs2_tpd_ocs} shows the $m/z$ 62 and $m/z$ 63 TPD curves corresponding to $^{18}$OCS and $^{18}$O$^{13}$CS in Exp. 1.
A desorption peak was detected at T $\sim$ 90 K, followed by a second peak at T $\sim$ 105 K. 
The $^{18}$OCS and $^{18}$O$^{13}$CS thermal desorption resulted in a decrease of the 2009 and 1955 cm$^{-1}$ IR band absorbances in the same temperature range, as observed in Fig. \ref{fig:co2cs2_7K_IR_evo}. 
The total OCS ice column density (calculated from the sum of $^{18}$OCS and $^{18}$O$^{13}$CS integrated IR absorbances) was $\sim$1.5 ML, 
%
%
representing $\sim$3\% of the initial sulfur in the ice, 
or $\sim$5\% of the sulfur consumed during irradiation. 

\begin{figure}
    \centering
    \includegraphics[width=0.8\linewidth]{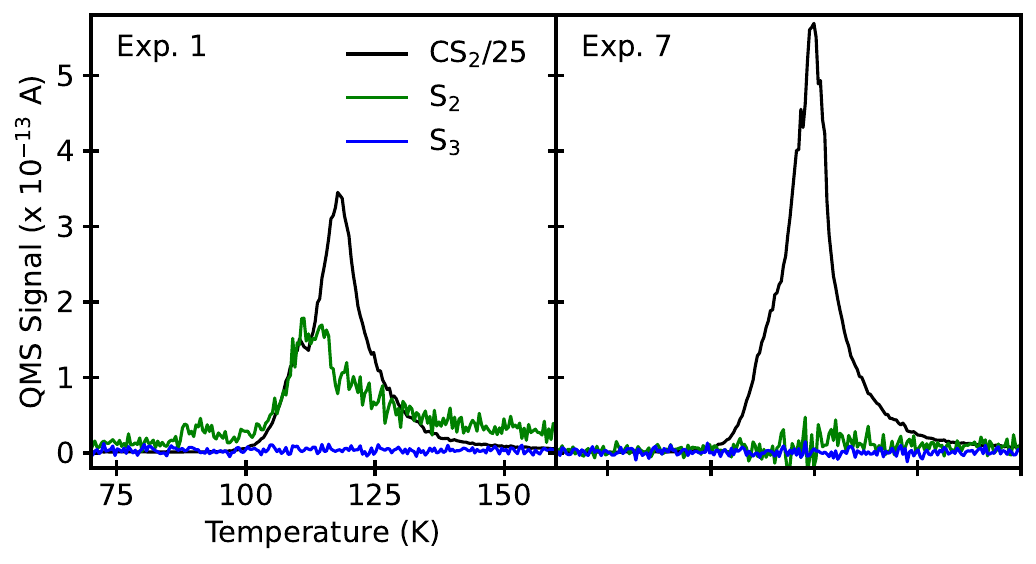}
    \caption{TPD curves corresponding to the remaining CS$_2$ (black), S$_2$ (green), and S$_3$ (blue) measured during thermal desorption of the irradiated $^{13}$C$^{18}$O$_2$:CS$_2$ ice sample in Exp. 1 (left panel), and $^{13}$C$^{18}$O:CS$_2$ ice sample in Exp. 7 (right panel).}
    \label{fig:tpd_s2}
\end{figure}

In addition to S$^{18}$O$_2$, S$^{18}$O$_3$, and both OCS isotopologs (accounting for $\sim$35\% of the sulfur consumed during irradiation), 
two additional S-bearing species were detected after irradiation of a $^{13}$C$^{18}$O$_2$:CS$_2$ ice sample: C$_3$S$_2$ and S$_2$. 
The IR feature detected at 2070 cm$^{-1}$ (Figures \ref{fig:co2cs2_IR} and \ref{fig:co2cs2_gauss}) was assigned to C$_3$S$_2$ based on \citet{maity13}. 
Unfortunately, no band strength was found in the literature for this feature.
%
%
As a first approximation, we estimated the C$_3$S$_2$ IR band strength from the C$_3$O$_2$ IR band strength in Table \ref{tab:ir}, assuming the same ratio as for the CS$_2$ and CO$_2$ main IR features. This led to a band strength of 1.8 $\times$ 10$^{-16}$ cm molecule$^{-1}$, and a C$_3$S$_2$ ice column density of $\sim$0.4 ML. 
%
%
On the other hand, S$_2$ was not detected in the IR spectrum due to its lack of dipole moment, but thermal desorption of this species was observed during the TPD (Fig. \ref{fig:tpd_s2}, left panel). 
S$_2$ molecules mainly desorbed in the 110$-$115 K temperature range, in agreement with the 113 K desorption temperature reported in \citet{cazaux22}. 
%
%
A rough estimation of the S$_2$ ice column density was derived from the integrated TPD curve using Eq. \ref{eqn_qms}. Using the electron-impact ionization cross-section listed in the NIST database (7.927 \AA$^2$), and assuming a fragmentation factor of 0.5 for S$_2$ as a first approximation,  
the estimated S$_2$ column density was $\sim$1 ML. 
%
Thermal desorption of S$_3$ molecules was not detected (Fig. \ref{fig:tpd_s2}, left panel), while the molecular mass of longer sulfur allotropes (S$_n$ with $n\ge$4) fell outside of the QMS mass range and could not be monitored.
As a result, approximately $\sim$60\% of the sulfur consumed during irradiation 
(corresponding to $\sim$35\% of the initial sulfur) 
could not be tracked at the end of Experiments 1 and 2. 
The nature of this missing sulfur is discussed in Sect. \ref{sec:disc_chem}.



Based on the reaction scheme proposed in \citet{maity13}, formation of $^{18}$OCS and $^{18}$O$^{13}$CS molecules in our experiments probably took place through the dissociation of $^{13}$C$^{18}$O$_2$ and CS$_2$ molecules, followed by the CS + $^{18}$O and $^{13}$C$^{18}$O + S atom-addition reactions. 
According to this scheme, the abundance of reactants available for both pathways would in principle be equivalent regardless of the number of dissociated $^{13}$C$^{18}$O$_2$ and CS$_2$ molecules during irradiation, because dissociation of these molecules would lead to the same amount of $^{13}$C$^{18}$O and $^{18}$O fragments, and CS and S fragments. 
%
In order to estimate the actual relative contribution of both pathways in our experiments, we 
calculated the area of the $^{18}$OCS and $^{18}$O$^{13}$CS Gaussians  
in the top left panel of Fig. \ref{fig:co2cs2_gauss}, 
obtaining a $^{18}$OCS/$^{18}$O$^{13}$CS ratio of $\sim$3.7 in Exp. 1 ($\sim$3.6 in Exp. 2). 
A similar ratio was obtained from the numerical integration of the $m/z$ 62 and $m/z$ 63 TPD curves shown in the bottom left panel of Fig. \ref{fig:co2cs2_tpd_ocs} ($\sim$2.4 in Exp. 1 and $\sim$2.3 in Exp. 2). 
%
Therefore, 
$\sim$75\% of the produced OCS molecules in our experiments were formed through the oxidation of CS (CS+O) pathway, while the remaining $\sim$25\% were formed through the sulfurization of CO (CO+S) pathway. This suggested that the CS + $^{18}$O $\rightarrow$ $^{18}$OCS atom-addition reaction was more favorable than the $^{13}$C$^{18}$O + S $\rightarrow$ $^{18}$O$^{13}$CS reaction.

\subsection{VUV photon irradiation of a $^{13}$C$^{18}$O$_2$:CS$_2$ ice sample at 10 K}\label{sec:co2_cs2_vuv}

In order to test whether the nature of the energetic processing could have an effect on the relative contributions of the CO+S and CS+O pathways to the formation of OCS, 
we irradiated a $^{13}$C$^{18}$O$_2$:CS$_2$ ice sample with VUV photons in Experiments 5 and 6. 
We note that the initial CS$_2$ ice abundance in Experiments 5 and 6 was $\sim$7\%, whereas that of Experiments 1 and 2 was $\sim$12$-$15\%. In addition, the ice irradiation temperature  was slightly higher ($\sim$10 K, the lowest temperature reached in the SPACE CAT experimental setup). In any case, we did not expect these differences to have a significant effect on the relative contributions of the two OCS formation pathways. 

\begin{figure}
    \centering
    \includegraphics[width=1\linewidth]{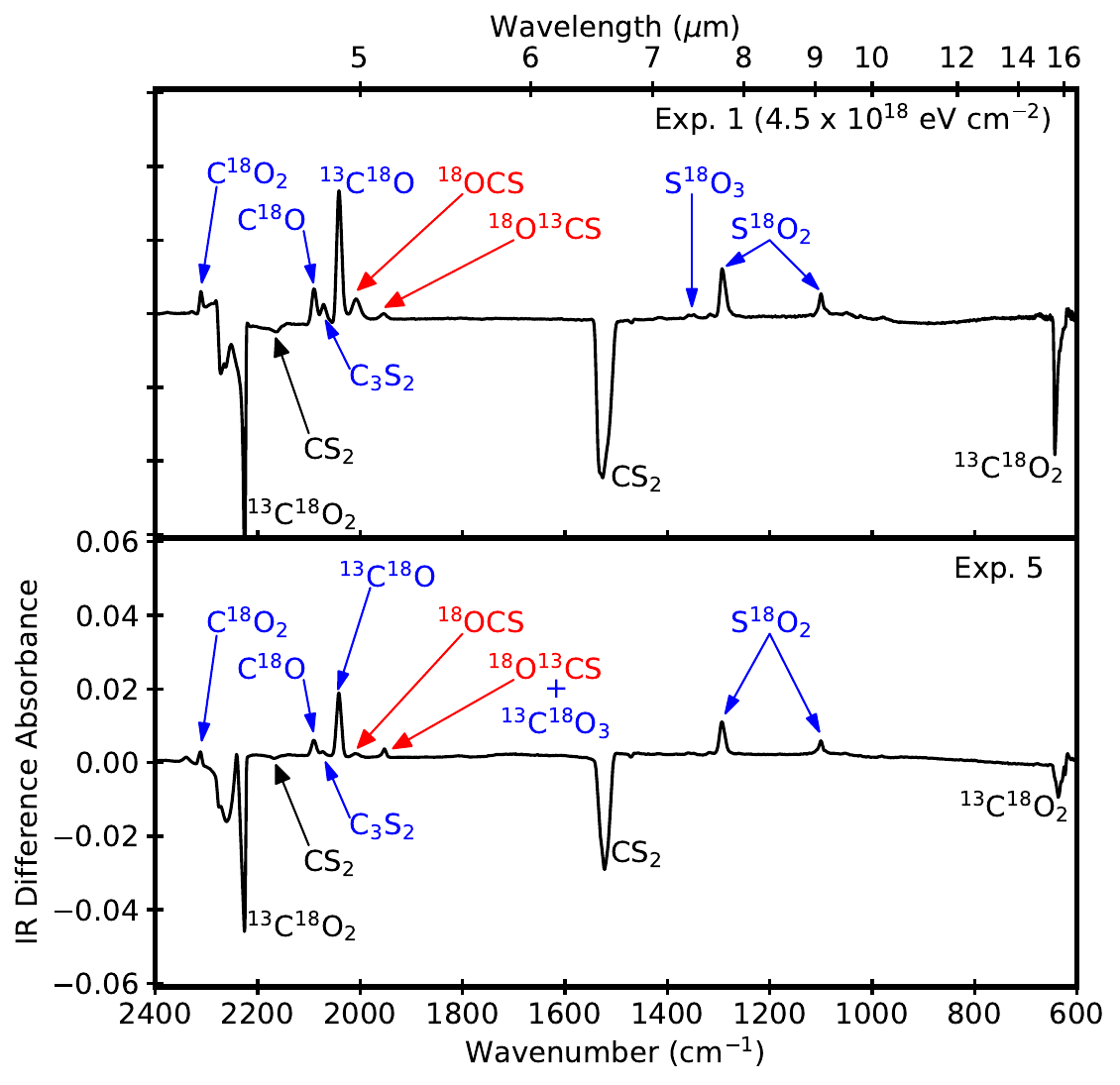}
    \caption{IR difference spectrum obtained upon irradiation of 4.5 $\times$ 10$^{18}$ eV cm$^{-2}$ with 2 keV electrons (top panel) and VUV photons (bottom panel) of a $^{13}$C$^{18}$O$_2$:CS$_2$ ice sample at 7$-$10 K in Experiments 1 and 5, respectively. IR band assignments are indicated for the initial ice components (black), and ice chemistry products (blue), including the OCS isotopologs (red).}
    \label{fig:co2cs2_e_vuv}
\end{figure}

\begin{figure}
    \centering
    \includegraphics[width=1\linewidth]{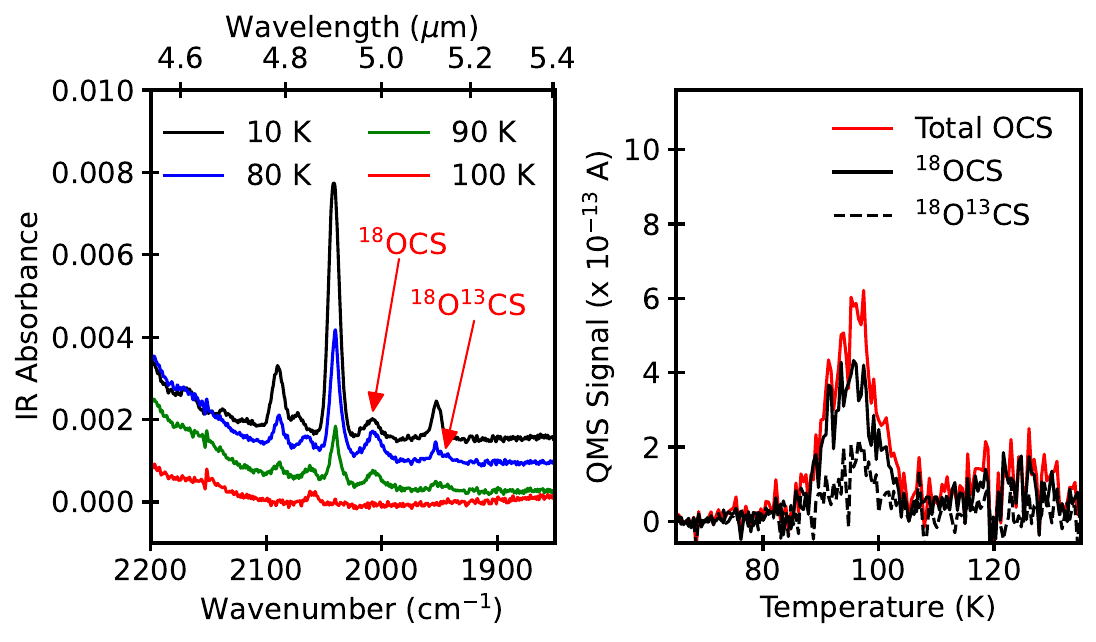}
    \caption{Left: evolution of the IR spectrum during the TPD of the VUV photon irradiated $^{13}$C$^{18}$O$_2$:CS$_2$ ice sample in Exp. 5. 
    The IR spectra during the TPD were collected with a 45$^\circ$ incidence angle. 
    Right: TPD curves corresponding to $^{18}$OCS (solid black), and $^{18}$O$^{13}$CS (dashed black), along with the sum of both signals (red) measured during thermal desorption of the VUV photon irradiated $^{13}$C$^{18}$O$_2$:CS$_2$ ice sample in Exp. 6. 
    Note that the ice sample was not facing the QMS during the TPD in Exp. 5 (so that IR spectra could be collected). This led to TPD curves with a lower signal-to-noise ratio. In any case, the TPD curves were similar in Experiments 5 and 6.}
    \label{fig:co2cs2_vuv_ir_tpd}
\end{figure}

The IR difference spectrum in the 2400$-$600 cm$^{-1}$ range after VUV photon irradiation of the ice sample in Exp. 5 is presented in the bottom panel of Fig. \ref{fig:co2cs2_e_vuv} (the results were similar in Exp. 6). 
For comparison, the top panel of Fig. \ref{fig:co2cs2_e_vuv} shows the IR difference spectrum of the 2 keV electron irradiated sample in Exp. 1 after a similar irradiated energy fluence. 
Qualitatively, both spectra looked very similar, but the chemistry seemed to proceed to a larger extent in the electron irradiated ice. 
Formation of both, $^{18}$OCS and $^{18}$O$^{13}$CS was also detected in the VUV irradiation experiment through the 2009 and 1955 cm$^{-1}$ IR features. 
However, in this experiment the 1955 cm$^{-1}$ feature presented a contribution from, at least, one additional molecule. 
While thermal desorption of $^{18}$OCS and $^{18}$O$^{13}$CS was detected above 85 K (Fig. \ref{fig:co2cs2_vuv_ir_tpd}, right panel), we observed a decrease in the absorbance of the 1955 cm$^{-1}$ feature between 10 and 80 K (Fig. \ref{fig:co2cs2_vuv_ir_tpd}, left panel). 
This indicated a contribution from a different species 
that desorbed or dissociated in the 10$-$80 K temperature range. 
We speculate that this species could be $^{13}$C$^{18}$O$_3$. As explained above, formation of CO$_3$ has been previously reported in VUV irradiation experiments of CO$_2$-containing ices \citep[see, e.g.,][]{martin15,asper15}, and the reported position of the corresponding IR feature (2044 cm$^{-1}$) overlapped with that of OCS \citep[2040 cm$^{-1}$,][]{asper15}. 
It is thus possible that the IR features of $^{13}$C$^{18}$O$_3$ and $^{18}$O$^{13}$CS also overlap. Unfortunately, thermal desorption of CO$_3$ was not detected in \citet{martin15}, and it was not observed either in Experiments 5 and 6, so we could not confirm the presence of this molecule. 

We followed the same procedure as in Sect. \ref{sec:co2_cs2_7K} to calculate the relative contribution of the two OCS formation pathways. To this purpose, we used the numerical integration of the $^{18}$OCS and $^{18}$O$^{13}$CS features in the 80 K IR spectrum (that did not present contributions from any other species), and the $m/z$ 62 and $m/z$ 63 TPD curves, leading to an average $^{18}$OCS/$^{18}$O$^{13}$CS ratio of $\sim$3.5. 
This indicated that $\sim$78\% of the produced OCS was formed through the CS+O pathway, and $\sim$22\% through the CO+S pathway, comparable to the values reported in Sect. \ref{sec:co2_cs2_7K} for the electron irradiation experiment. 
Therefore, the relative contribution of the two OCS formation pathways upon irradiation of $^{13}$C$^{18}$O$_2$:CS$_2$ ice samples did not significantly depend on the nature of the energetic processing. 
%

\subsection{2 keV electron irradiation of a $^{13}$C$^{18}$O:CS$_2$ ice sample at 7 K}\label{sec:co_cs2_7K}

\begin{figure}
    \centering
    \includegraphics[width=1\linewidth]{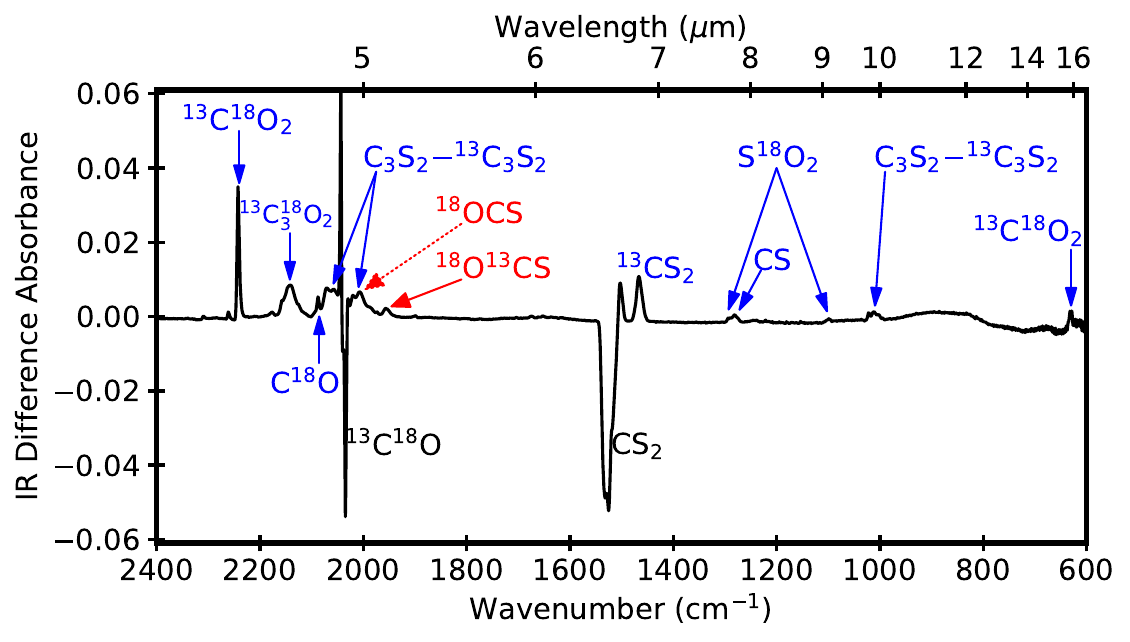}
    \caption{IR difference spectra obtained upon 2 keV electron irradiation of a $^{13}$C$^{18}$O:CS$_2$ ice sample at 7 K (Exp. 7).  
    IR band assignments are indicated for the initial ice components (black), and ice chemistry products (blue), including the OCS isotopologs (red).}
    \label{fig:cocs2_IR}
\end{figure}

\begin{figure*}
    \centering
    \includegraphics[width=0.61\linewidth]{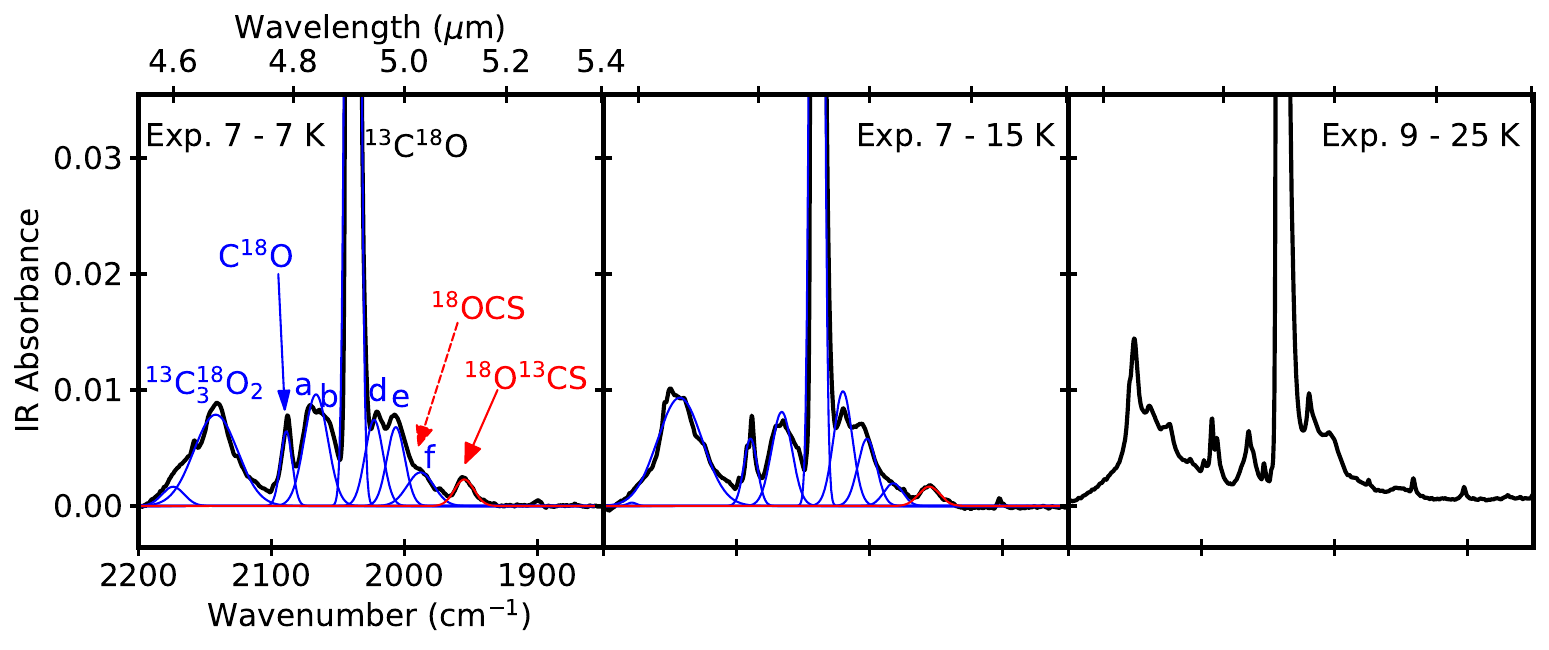}
    \includegraphics[width=0.6\linewidth]{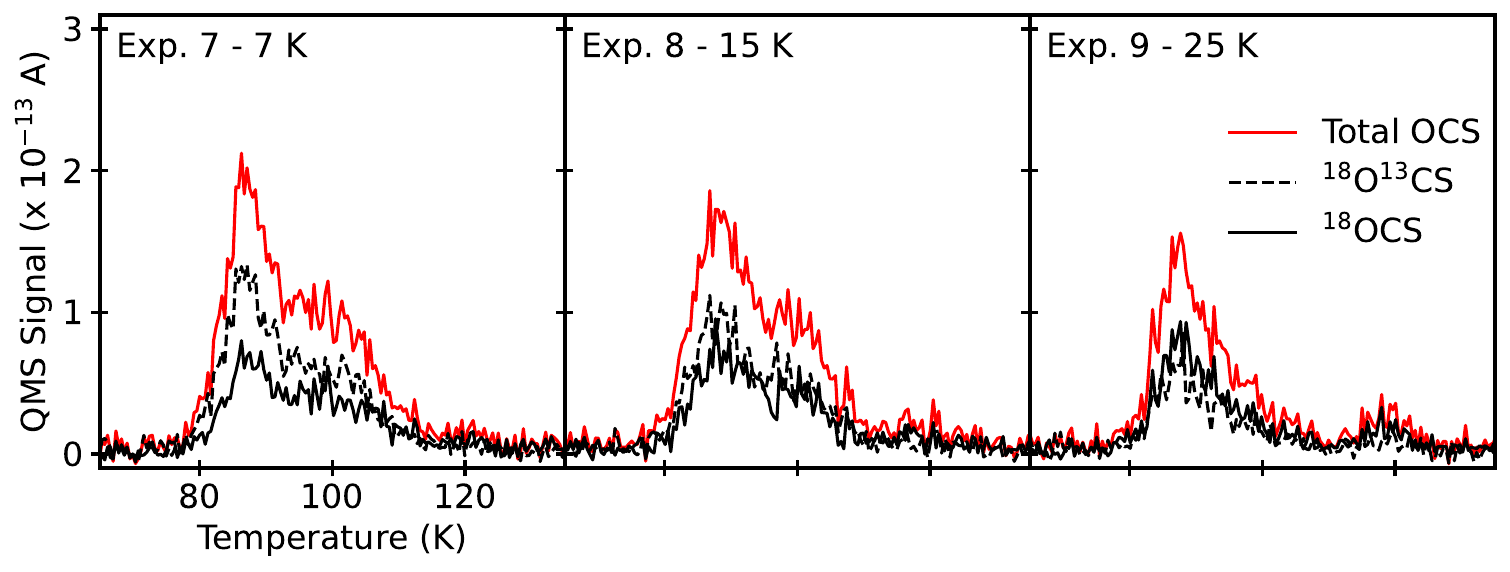}
    \caption{\textit{Top}: Gaussian fitting of the features in the IR spectra of the $^{13}$C$^{18}$O:CS$_2$ ice samples irradiated at 7 K (left panel) and 15 K (middle panel). The spectra of the experiment at 25 K could not be fitted (right panel). IR band assignments are indicated for the initial ice components (black), and ice chemistry products (blue), including $^{18}$O$^{13}$CS (red). The expected position of the $^{18}$OCS feature is indicated, but it could not be detected due to overlapping with features corresponding to C$_3$S$_2$ isotopologs. The positions of the latter are indicated with letters a$-$f, corresponding to a) C$_3$S$_2$, b) SCC$^{13}$CS (these two features were fitted with a single Gaussian), d) SC$^{13}$CCS, e)S$^{13}$C$^{13}$CCS, and f) $^{13}$C$_3$S$_2$. The S$^{13}$CC$^{13}$CS feature overlapped with the $^{13}$C$^{18}$O band and could not be detected either. 
    \textit{Bottom}: TPD curves corresponding to $^{18}$OCS (solid black) and $^{18}$O$^{13}$CS (dashed black), along with the sum of both signals (solid red), measured during thermal desorption of $^{13}$C$^{18}$O:CS$_2$ ice samples irradiated at 7 K (left panel), 15 K (middle panel), and 25 K (right panel). 
    }
    \label{fig:cocs2_tpd_ocs}
    \label{fig:cocs2_gauss}
\end{figure*}

Figure \ref{fig:cocs2_IR} shows the IR difference spectrum in the 2400$-$600 cm$^{-1}$ range obtained upon 2 keV electron irradiation at 7 K of the $^{13}$C$^{18}$O:CS$_2$ ice sample in Exp. 7.  
In this experiment, we observed a lower depletion of CS$_2$ molecules  
($\sim$40\% in Exp. 7, 
compared to $\sim$60\% in Experiments 1 and 2),  
even though the ice thickness, initial CS$_2$ ice abundance, and irradiated energy fluence were similar to those reported in Sect. \ref{sec:co2_cs2_7K}. 
Therefore, sulfur chemistry proceeded to a lower extent in the irradiated $^{13}$C$^{18}$O:CS$_2$ sample compared to a $^{13}$C$^{18}$O$_2$:CS$_2$ ice. 
%

Irradiation of the $^{13}$C$^{18}$O:CS$_2$ ice sample led to the formation of  $^{13}$C$^{18}$O$_2$ (with two IR features detected at 2242 and 630 cm$^{-1}$) and $^{13}$C$_3$$^{18}$O$_2$ (with a broad IR feature detected at $\sim$2140 cm$^{-1}$).  
The assignment of the $\sim$2140 cm$^{-1}$ feature to $^{13}$C$_3$$^{18}$O$_2$ was based on a preliminary 2 keV electron irradiation experiment of a CO:CS$_2$ ice mixture (not shown in this paper) in which the C$_3$O$_2$ feature was detected at 2242 cm$^{-1}$ \citep{sicilia12}. 
In addition, formation of C$^{18}$O was also observed at 2090 cm$^{-1}$, as in the $^{13}$C$^{18}$O$_2$:CS$_2$ ice samples. 
%
%

Regarding the S-bearing products, S$^{18}$O$_2$ and S$^{18}$O$_3$ formation proceeded to a much lower extent in the $^{13}$C$^{18}$O:CS$_2$ ice mixture. The integrated absorbance of the S$^{18}$O$_2$ 1290 cm$^{-1}$ feature was a factor of $\sim$50 lower compared to the irradiated $^{13}$C$^{18}$O$_2$:CS$_2$ sample (corresponding to $\sim$0.2 ML), 
while the IR feature corresponding to S$^{18}$O$_3$ was not detected. 
The signal of the S$^{18}$O$_2$ TPD curve was also between one and two orders of magnitude lower (see Fig. \ref{fig:cocs2_tpd_so2} in Appendix \ref{app:co_cs2}). 
%
This could stem from the lower abundance of $^{18}$O atoms in a $^{13}$C$^{18}$O:CS$_2$ ice with the same molecular composition than a $^{13}$C$^{18}$O$_2$:CS$_2$ sample (see Table \ref{tab:exp}). 
A consequence of the lower absorbance of the 1290 cm$^{-1}$ band was the identification of a small CS IR feature at 1280 cm$^{-1}$ \citep{bohn92,bahou00}. This feature could also be present in the spectrum of the irradiated $^{13}$C$^{18}$O$_2$:CS$_2$ sample, but hindered by the more intense S$^{18}$O$_2$ band. 
As for S$^{18}$O$_2$ and S$^{18}$O$_3$, formation of S$_2$ was also hampered in this experiment, and this molecule was not detected during the TPD of the irradiated ice (right panel of Fig. \ref{fig:tpd_s2}). 

OCS molecules were formed through both the CS+O and the CO+S pathways also in the $^{13}$C$^{18}$O:CS$_2$ ice sample. 
Unfortunately, the 2009 cm$^{-1}$ $^{18}$OCS IR feature overlapped with C$_3$S$_2$ isotopolog bands (see below), and   
only the 1955 cm$^{-1}$ $^{18}$O$^{13}$CS feature could be unambiguously identified   
in the IR spectrum 
The integrated absorbance of the latter was similar to that measured after irradiation of a $^{13}$C$^{18}$O$_2$:CS$_2$ ice sample. 
In any case, thermal desorption of both molecules was detected during the TPD of the irradiated ice (Fig. \ref{fig:cocs2_tpd_ocs}, bottom left panel). 
According to the sum of the $m/z$ 62 and $m/z$ 63 QMS signals, the total number of produced OCS molecules was a factor of $\sim$2.5 lower than in the $^{13}$C$^{18}$O$_2$:CS$_2$ experiments, for a total of $\sim$0.6 ML of OCS. 

On the other hand, formation of C$_3$S$_2$ proceeded to a higher extent in the irradiated  $^{13}$C$^{18}$O:CS$_2$ ice sample, and was the main S-bearing irradiation product in this experiment. 
The top left panel of Fig. \ref{fig:cocs2_gauss} shows a multiple Gaussian fitting of the 2220$-$1850 cm$^{-1}$ region of the spectrum. 
Multiple features with similar absorbances were detected at 2070 (feature a in the Figure), 2055 (b), 2020 (d), 2007 (e), and 1985 cm$^{-1}$ (f), corresponding to C=C antisymmetric stretching of C$_3$S$_2$ and the isotopologs SCC$^{13}$CS, SC$^{13}$CCS, S$^{13}$C$^{13}$CCS, and $^{13}$C$_3$S$_2$, respectively \citep{bohn92}. 
Only the feature corresponding to S$^{13}$CC$^{13}$CS was missing, because it overlapped with that of $^{13}$C$^{18}$O.
%
Four additional features were detected at 1022, 1012, 1002, and 995 cm$^{-1}$, corresponding to the C=S antisymmetric stretching of the same molecules \citep{bohn92}. 
In addition, thermal desorption of these molecules was detected above 150 K through the C$_2$S molecular fragments (see bottom left panel of Fig. \ref{fig:cocs2_tpd_c3s2} in Appendix \ref{app:co_cs2}). 
%
%
The sum of integrated absorbances of the C=C antisymmetric stretching features was $\sim$10 times higher%
\footnote{Note that for this calculation we assumed, as a first approximation, that the integrated absorbance corresponding to the S$^{13}$CC$^{13}$CS isotopolog (whose feature overlapped with the $^{13}$C$^{18}$O band) was the same as that of the S$^{13}$C$^{13}$CCS isotopolog (feature e in the top left panel of Fig. \ref{fig:cocs2_gauss}).  In addition, we subtracted the expected contribution of the $^{18}$OCS IR feature from the sum of C$_3$S$_2$ isotopolog integrated absorbances. This contribution was estimated from the $^{18}$O$^{13}$CS integrated absorbance and the ($m/z$ 62)/($m/z$ 63) QMS signal ratio.}
than after irradiation of a $^{13}$C$^{18}$O$_2$:CS$_2$ ice sample in Experiments 1 and 2. 
The estimated column density of C$_3$S$_2$ and the different isotopologs was $\sim$3.5 ML. 
%
The detection of multiple C$_3$S$_2$ isotopologs with similar abundances in this experiment contrasted with the formation of only C$_3$S$_2$ in the irradiated $^{13}$C$^{18}$O$_2$:CS$_2$ samples, and suggested that $^{13}$C atoms coming from $^{13}$C$^{18}$O played a significant role in the formation of C$_3$S$_2$ in this experiment. 
The interaction of $^{13}$C with S was also evidenced by the formation of $\sim$1.5 ML of $^{13}$CS$_2$ molecules (detected at 1468 cm$^{-1}$), that was not observed in Experiments 1 and 2. 
%
After including $\sim$7 ML of S contained in C$_3$S$_2$ isotopologs, 
$\sim$3 ML contained in $^{13}$CS$_2$, 
$\sim$0.6 ML contained in both OCS isotopologs, 
and $\sim$0.2 ML in S$^{18}$O$_2$ molecules,  
roughly $\sim$45\% of the consumed sulfur, corresponding to $\sim$19\% of the initial sulfur, was missing at the end of the experiment. 


Regarding the contribution of the CO+S and CS+O pathways, only $^{18}$O$^{13}$CS could be unambiguously identified in the IR spectrum, but thermal desorption of both, $^{18}$O$^{13}$CS and $^{18}$OCS, was detected (as explained above).  
Therefore, we used the corresponding TPD curves to evaluate the relative contribution of the two OCS formation pathways. 
While the ice abundance of $^{18}$O$^{13}$CS 
was similar in experiments with a CO$_2$- and a CO-ice matrix, 
the TPD curve area corresponding to the $^{18}$OCS molecules 
decreased by a factor of $\sim$4 in the irradiated $^{13}$C$^{18}$O:CS$_2$ ice compared to a $^{13}$C$^{18}$O$_2$:CS$_2$ sample.
%
As for S$^{18}$O$_2$, the lower $^{18}$OCS formation could be due to the lower availability of $^{18}$O atoms. 
According to the integrated $m/z$ 62 and $m/z$ 63 TPD curves, the $^{18}$OCS/$^{18}$O$^{13}$CS ratio was $\sim$0.7.   
This means that $\sim$40\% of the OCS molecules were formed through the CS+O pathway 
and $\sim$60\% were formed through the CO+S pathway. 
%
The higher contribution of the CO+S pathway in the $^{13}$C$^{18}$O:CS$_2$ ice sample was probably the result of the higher abundance of $^{13}$C$^{18}$O compared to any other of the required reactants for the formation of OCS (i.e., CS, S, $^{18}$O).

\subsection{2 keV electron irradiation of $^{13}$C$^{18}$O$_2$:CS$_2$ ice samples at 25 K and 50 K}\label{sec:co2cs2_T}

\begin{figure*}
    \centering
    \includegraphics[width=0.6\linewidth]{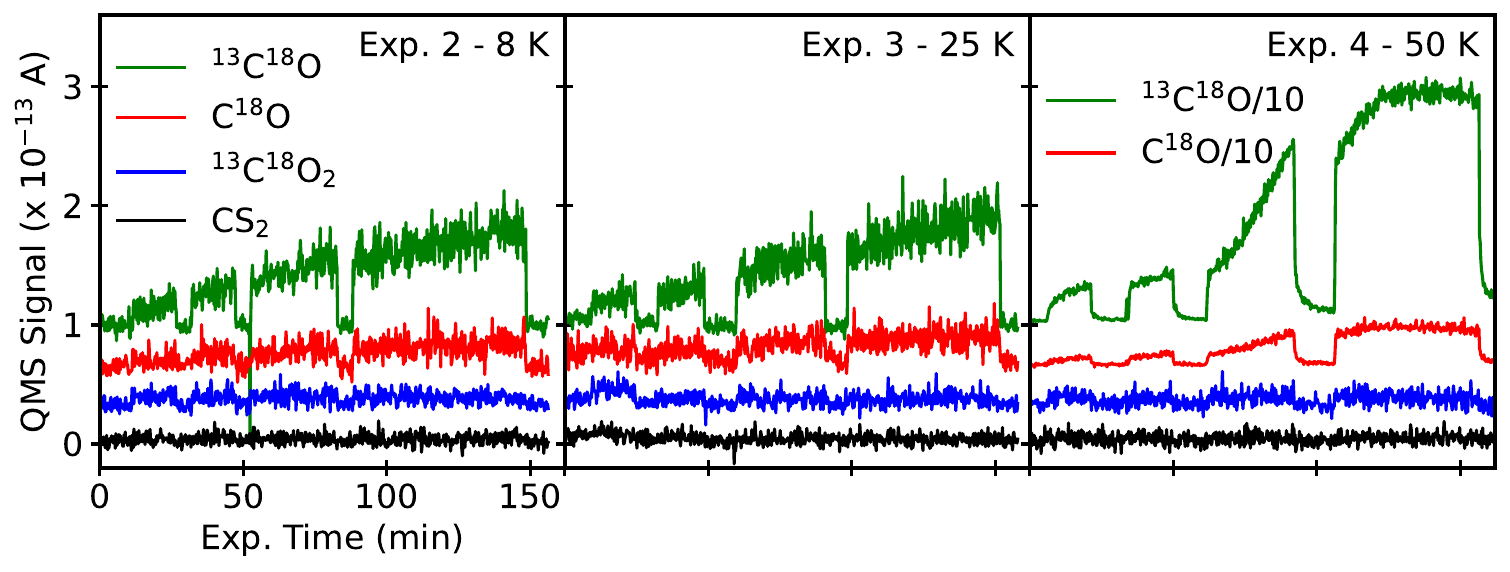}
    \caption{QMS signals corresponding to $^{13}$C$^{18}$O$_2$ (blue), CS$_2$ (black), $^{13}$C$^{18}$O (green), and C$^{18}$O (red) during irradiation of $^{13}$C$^{18}$O$_2$:CS$_2$ ice samples at 8 K (left panel), 25 K (middle panel) and 50 K (right panel). 
    The signals are offset for clarity. 
    The observed valleys correspond to pauses in the irradiation of the samples. 
    We note that the QMS signals were not monitored during irradiation in Exp. 1.}
    \label{fig:co2cs2_irr_qms}
\end{figure*}

The IR difference spectra of the $^{13}$C$^{18}$O$_2$:CS$_2$ ice samples irradiated at 25 K and 50 K (Experiments 3 and 4) were similar to those of the samples irradiated at 7 K, and are shown in Fig. \ref{fig:co2_cs2_IR_allT} of the Appendix \ref{app:co2_cs2}). 
No significant differences were found in the depletion of $^{13}$C$^{18}$O$_2$ and CS$_2$ molecules upon irradiation. 
%
The main irradiation product was $^{13}$C$^{18}$O in all experiments. 
However, the corresponding IR integrated absorbance was $\sim$35\% lower (corresponding to $\sim$20 ML less) in the ice sample irradiated at 50 K (Fig. \ref{fig:co2cs2_gauss}, top right panel). 
A $\sim$55\% decrease was also observed for C$^{18}$O.  
%
%
This could be due to a higher desorption of these molecules during irradiation in the 50 K experiment. 
Desorption of $^{13}$C$^{18}$O and C$^{18}$O molecules (along with a smaller amount of $^{13}$C$^{18}$O$_2$ molecules) was observed during irradiation of the $^{13}$C$^{18}$O$_2$:CS$_2$ samples at any temperature (Fig. \ref{fig:co2cs2_irr_qms}). 
Using the calibration of the QMS described in Sect. \ref{sec:exp_TPD} (as a first approximation, see Appendix \ref{app:kco} for more information), we estimated that $\sim$1 ML of $^{13}$C$^{18}$O desorbed during irradiation at 7 K and 25 K, 
while $\sim$20 ML desorbed at 50 K. 
This suggests that the formation of $^{13}$C$^{18}$O proceeded to the same extent in all experiments, but a higher fraction of the formed molecules desorbed during irradiation in the experiment at 50 K. 
Similar results were obtained for the C$^{18}$O desorption. 

Regarding the S-bearing products, formation of S$^{18}$O$_2$, S$^{18}$O$_3$, and S$_2$ proceeded to a similar extent in Experiments 1$-$4 (see Figures \ref{fig:co2_cs2_IR_allT} and \ref{fig:co2cs2_tpd_so2} in Appendix \ref{app:co2_cs2}). 
%
On the other hand, the C$_3$S$_2$ IR absorbance was $\sim$40\% lower in the sample irradiated at 50 K (Fig. \ref{fig:co2cs2_gauss}, top right panel). 
We speculate that this decrease could be related to the slight increase in the formation of $^{18}$OCS molecules through the CS + $^{18}$O reaction at  50 K (see below), leading to a lower availability of CS for the formation of C$_3$S$_2$. 

\begin{figure}
    \centering
    \includegraphics[width=0.75\linewidth]{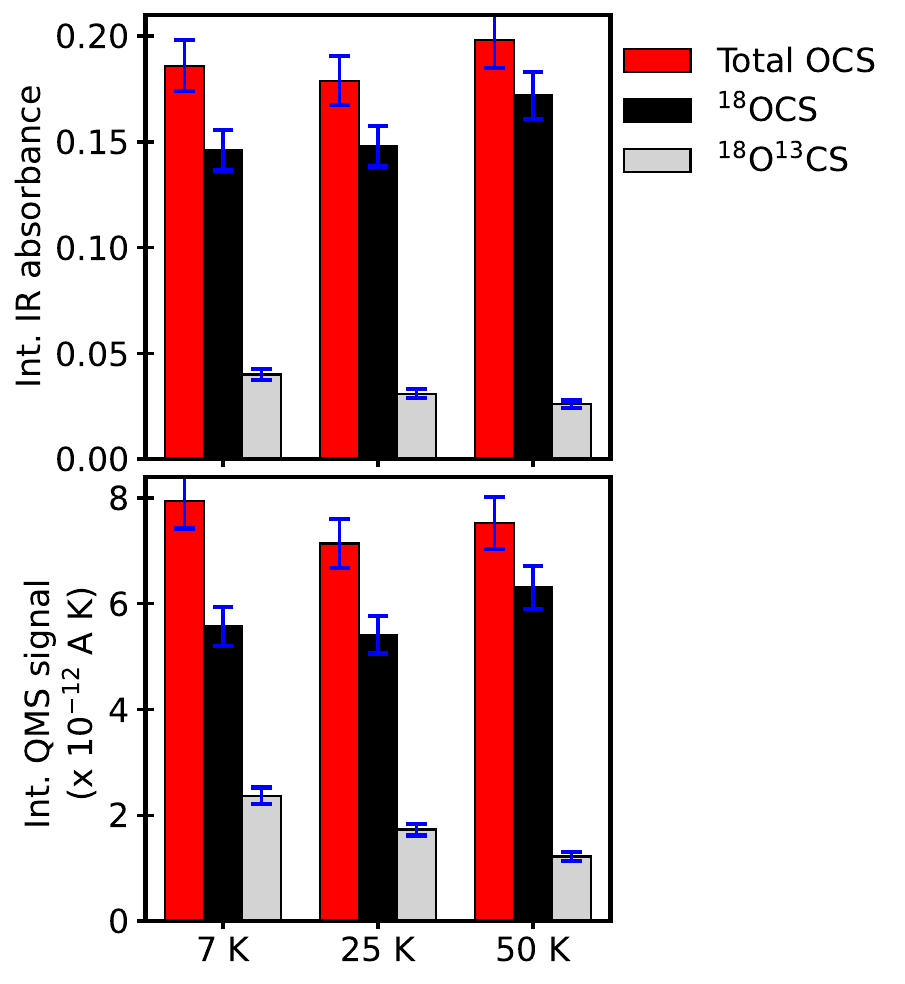}
    \caption{Formation of $^{18}$OCS (black) and $^{18}$O$^{13}$CS (gray), along with the total formation of OCS molecules (red) upon irradiation of $^{13}$C$^{18}$O$_2$:CS$_2$ ice samples at 7$-$50 K, measured as the integration of the corresponding IR features (top panel) and TPD curves (bottom panel).
    The values at 7 K are the average of Experiments 1 and 2. A 5\% relative experimental uncertainty (blue error bars) was assumed from the differences found between Experiments 1 and 2 (performed under similar conditions).
    }
    \label{fig:formation_ocs_co2_cs2}
\end{figure}

The total formation of OCS molecules did not present significant differences in the 7$-$50 K temperature range. 
However, the relative contribution of the CS+O and CO+S pathways slightly changed with the irradiation temperature, according to the corresponding IR features (Fig. \ref{fig:co2cs2_gauss}, top panels) and TPD curves (Fig. \ref{fig:co2cs2_gauss}, bottom panels). 
The integrated IR absorbances (top panel) and TPD curves (bottom panel) of $^{18}$OCS and $^{18}$O$^{13}$CS (along with the sum of both isotopologs), measured after irradiation of the ice samples at 7, 25, and 50 K, are presented in Fig. \ref{fig:formation_ocs_co2_cs2}, . 
The contribution of the CS+O pathway (leading to the formation of $^{18}$OCS molecules) increased from $\sim$75\% at 7 K (Sect. \ref{sec:co2_cs2_7K}) to $\sim$85\% at 50 K. 
%
%
%
As observed in Fig. \ref{fig:formation_ocs_co2_cs2}, 
this increase was mainly due to a gradual decrease in the $^{18}$O$^{13}$CS formation through the $^{13}$C$^{18}$O + S reaction, combined with a small increase in the $^{18}$OCS formation at 50 K. 
We note that the decrease in the $^{18}$O$^{13}$CS formation could not be due only to the lower $^{13}$C$^{18}$O ice column density measured in the 50 K experiment, since this decrease was already observed at 25 K (Fig. \ref{fig:formation_ocs_co2_cs2}).   
Therefore, the lower extent of the $^{13}$C$^{18}$O + S reaction as the temperature increased should have an additional explanation. 
This is discussed in Sect. \ref{sec:disc_ocsT}. 

\subsection{2 keV electron irradiation of $^{13}$C$^{18}$O:CS$_2$ ice samples at 15 K and 25 K}\label{sec:cocs2_T}

The IR difference spectra of the $^{13}$C$^{18}$O:CS$_2$ ice samples irradiated at 15 K and 25 K (Experiments 8 and 9) are shown in Appendix \ref{app:co_cs2} (Fig. \ref{fig:cocs2_IR_allT}, middle and bottom panels, respectively). 
The CS$_2$ depletion upon irradiation decreased from $\sim$40\% at 7 and 15 K to $\sim$20\% at 25 K. 
%
This did not affect the formation of $^{13}$C$^{18}$O$_2$ and $^{13}$C$_3^{18}$O$_2$, that did not present any significant differences in the 7$-$25 K temperature range according to their IR absorbances. 
%
Regarding the S-bearing products, S$^{18}$O$_2$ formation did not seem to drastically change with temperature either (see also Fig. \ref{fig:cocs2_tpd_so2} in Appendix \ref{app:co_cs2}).  
On the other hand, the formation of C$_3$S$_2$ and its isotopologs decreased by a factor of $\sim$2 in the experiment at 25 K, according to the numerical integration of the $\sim$1010 cm$^{-1}$ IR features and TPD curves (Fig. \ref{fig:cocs2_tpd_c3s2} in Appendix \ref{app:co_cs2}). 
The same decrease was observed for the formation of $^{13}$CS$_2$. 

\begin{figure}
    \centering
    \includegraphics[width=0.75\linewidth]{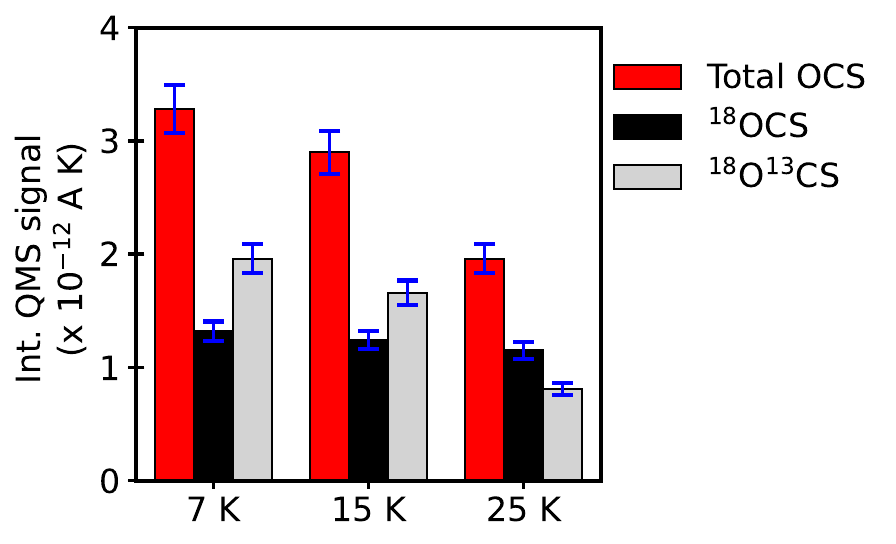}
    \caption{Formation of $^{18}$OCS (black) and $^{18}$O$^{13}$CS (gray), along with the total formation of OCS molecules (red) upon irradiation of $^{13}$C$^{18}$O:CS$_2$ ice samples at 7$-$25 K, measured as the integration of the corresponding TPD curves.
    The assumed 5\% relative experimental uncertainty is indicated as blue error bars.}
    \label{fig:formation_ocs_co_cs2}
\end{figure}

The effect of the irradiation temperature on the CS+O and CO+S OCS formation pathways was similar to that observed in the $^{13}$C$^{18}$O$_2$:CS$_2$ mixtures (Sect. \ref{sec:co2cs2_T}). 
The integrated TPD curves corresponding to $^{18}$OCS and $^{18}$O$^{13}$CS
(along with the sum of both isotopologs) in the experiments at 7, 15, and 25 K are shown in Fig. \ref{fig:formation_ocs_co_cs2}.
Formation of $^{18}$OCS through the CS + $^{18}$O reaction did not significantly change with temperature (even though dissociation of CS$_2$ proceeded to a lower extent at 25 K, as mentioned above). 
At the same time, the formation of $^{18}$O$^{13}$CS through the $^{13}$C$^{18}$O + S pathway gradually decreased with temperature. 
As a result, the relative contribution of the CS + $^{18}$O pathway to the total formation of OCS increased from $\sim$40\% at 7 K to $\sim$60\% at 25 K. 
The observed decrease from 7 to 25 K in the $^{13}$C$^{18}$O + S  $\rightarrow$ $^{18}$O$^{13}$CS reaction was higher in the $^{13}$C$^{18}$O:CS$_2$ samples (a factor of $\sim$2.5) than in the $^{13}$C$^{18}$O$_2$:CS$_2$ experiments between the same temperatures ($\sim$25\%). 
As in the latter experiments, this decrease was probably not due to a lower availability of $^{13}$C$^{18}$O molecules in the ice at higher temperatures, since formation of $^{13}$C$^{18}$O$_2$ and $^{13}$C$_3^{18}$O$_2$ proceeded to the same extent at all temperatures (as explained above). 
Likewise, the lower dissociation of CS$_2$ molecules at 25 K (leading to a decrease in the availability of CS and S) could not be the only responsible either, since the decrease was already observed at 15 K (with the same CS$_2$ dissociation as at 7 K). 

\subsection{2 keV electron irradiation of H$_2^{18}$O:CS$_2$ ice samples at 8 K and 50 K}\label{sec:h2o_cs2}

\begin{figure}
    \centering
    \includegraphics[width=1\linewidth]{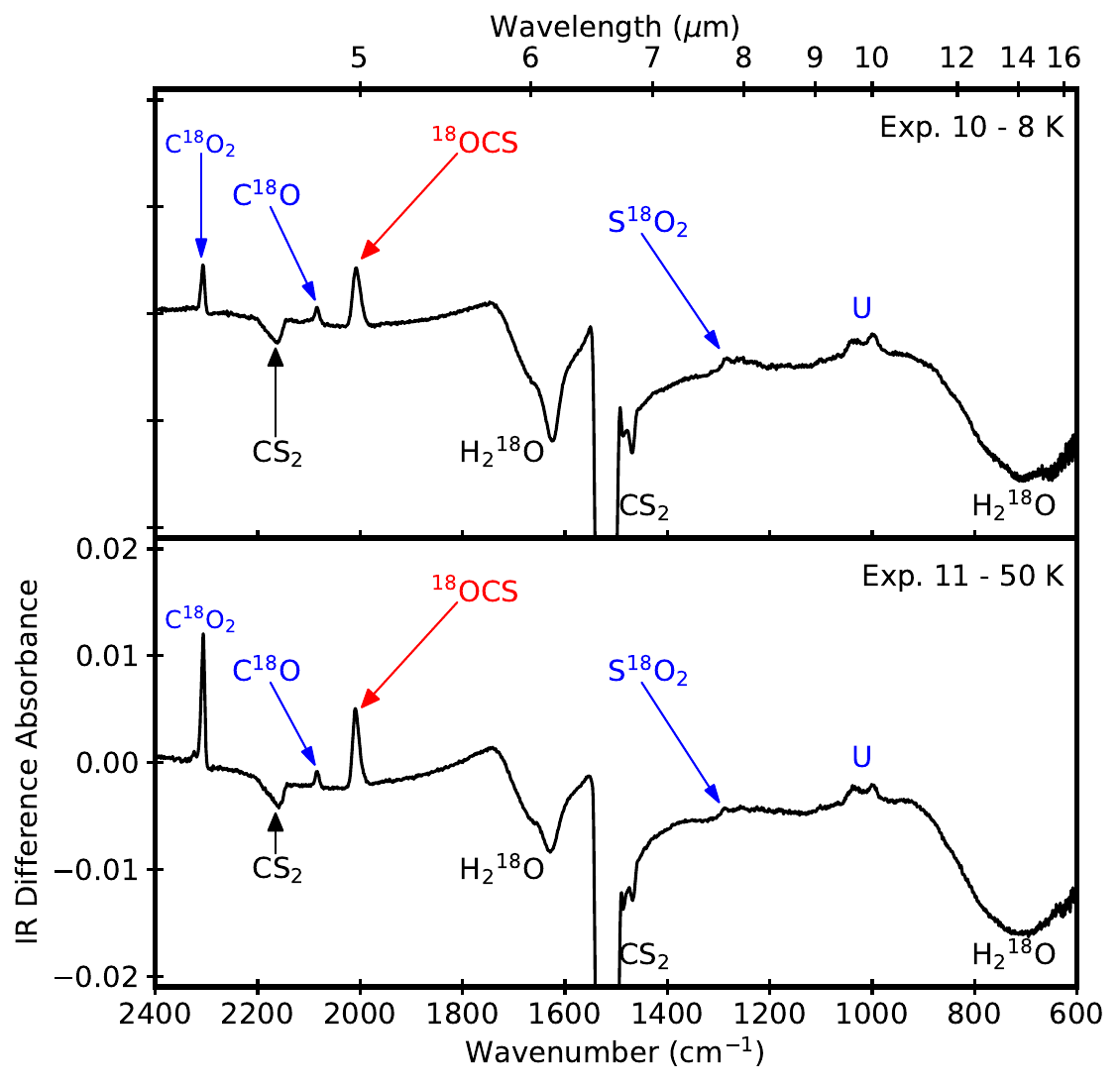}
    \caption{IR difference spectra obtained upon 2 keV electron irradiation of a H$_2^{18}$O:CS$_2$ ice sample at 8 K (Exp. 10, top panel) and 50 K (Exp. 11, bottom panel).
    IR band assignments are indicated for the initial ice components (black), and ice chemistry products (blue), including $^{18}$OCS (red).
    The unidentified features (U) could be related to structural changes in the H$_2^{18}$O ice matrix.  
    Further experiments beyond the scope of this paper would be required to confirm this scenario.}
    \label{fig:h2ocs2_ir}
\end{figure}

\begin{figure}
    \centering
    \includegraphics[width=1\linewidth]{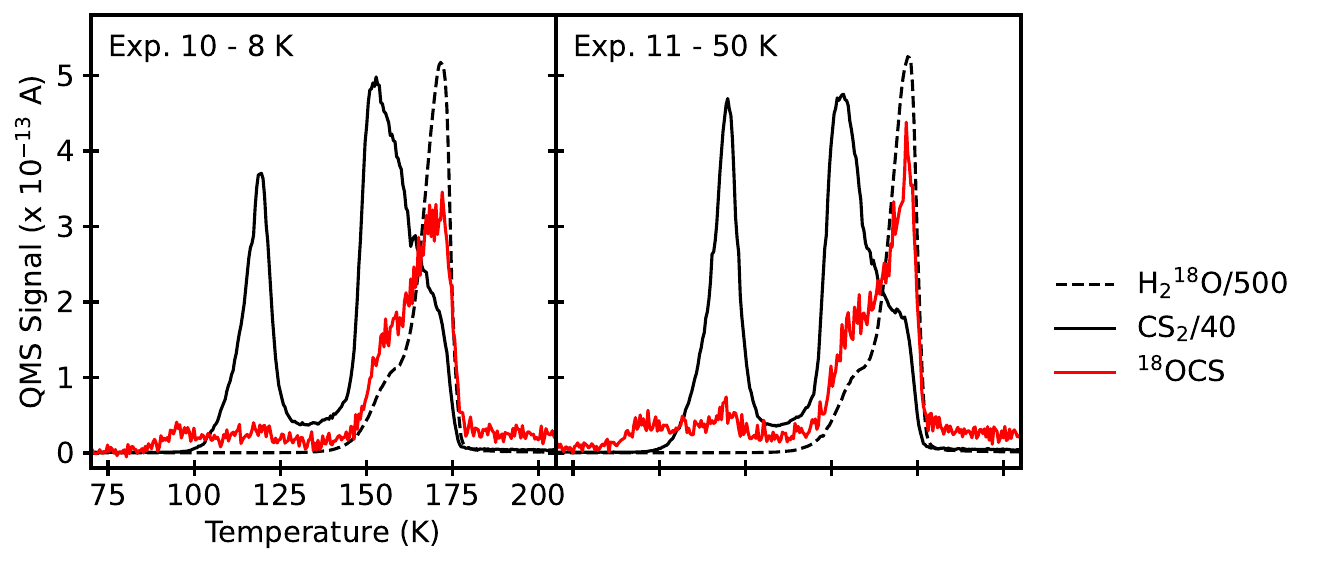}
    \caption{TPD curves corresponding to the remaining H$_2^{18}$O (dashed black) and CS$_2$ (solid black), and the produced $^{18}$OCS (solid red), measured during thermal desorption of the H$_2^{18}$O:CS$_2$ ice samples irradiated at 8 K (Exp. 10, left panel) and 50 K (Exp. 11, right panel). 
    }
    \label{fig:tpd_h2o_cs2}
\end{figure}

In order to explore the CS+O pathway for OCS formation and its dependence with temperature in a different ice environment, we performed 2 keV electron irradiation experiments of H$_2^{18}$O:CS$_2$ ice samples at 8 and 50 K (Experiments 10 and 11 in Table \ref{tab:exp}).  
In these experiments CS$_2$ molecules acted as a source of CS and H$_2^{18}$O as a source of $^{18}$O atoms. 
Therefore, OCS molecules most likely formed through a CS+O pathway, either through the CS + $^{18}$O atom-addition reaction proposed in \citet{maity13} (Eq. \ref{eq:cs+o}), and/or through the CS + OH neutral-neutral reaction described in \citet{adriaens10} (see Sect. \ref{sec:intro}). 

Fig. \ref{fig:h2ocs2_ir} shows the IR difference spectra in the 2400$-$600 cm$^{-1}$ range for the irradiation at 8 K (top panel) and 50 K (bottom panel). 
In both experiments, $\sim$15\% of the initial H$_2^{18}$O 
and $\sim$22\% of the initial CS$_2$ molecules 
were depleted upon irradiation (according to the $\sim$780 cm$^{-1}$ and $\sim$1500 cm$^{-1}$ IR features, respectively). 
The IR difference spectra revealed the formation of C$^{18}$O and C$^{18}$O$_2$, with IR features detected at 2090 cm$^{-1}$ and 2310 cm$^{-1}$ (respectively). 
As in the $^{13}$C$^{18}$O$_2$:CS$_2$ and $^{13}$C$^{18}$O:CS$_2$ experiments (Sect. \ref{sec:co2_cs2_7K}), formation of C$^{18}$O could have taken place through addition reactions to C atoms from completely dissociated CS$_2$ molecules, and/or through dissociation of previously produced $^{18}$OCS molecules. 
%

On the other hand, three S-bearing products were detected upon irradiation of the H$_2^{18}$O:CS$_2$ ice samples. 
Formation of S$^{18}$O$_2$ was detected in the IR difference spectra at 1290 cm$^{-1}$ (Fig. \ref{fig:h2ocs2_ir}), and during the TPD of the irradiated ice samples (Fig. \ref{fig:tpd_h2o_cs2_app} in Appendix \ref{app:h2o_cs2}).
The estimated ice column density was $\sim$0.2 ML at 8 K, and a factor of 2 lower at 50 K, representing less than 1\% of the consumed S during irradiation. 
We also tentatively detected co-desorption of less than 1 ML of S$_2$ molecules with H$_2^{18}$O during the TPD of the irradiated ice samples (Fig. \ref{fig:tpd_h2o_cs2_app} in Appendix \ref{app:h2o_cs2}). 
%
%
%
The main detected S-bearing product was $^{18}$OCS, with an IR feature at 2009 cm$^{-1}$  (Fig. \ref{fig:h2ocs2_ir}). 
The corresponding TPD curves 
are shown in Fig. \ref{fig:tpd_h2o_cs2}. 
We note that while a small fraction of the $^{18}$OCS molecules desorbed at $\sim$95 K or co-desorbed with CS$_2$ at $\sim$120 K, most of the produced $^{18}$OCS in these experiments was entrapped in the H$_2^{18}$O ice matrix and co-desorbed with this species at $\sim$170 K. 
According to the integrated IR absorbance, the $^{18}$OCS ice column density was $\sim$1 ML in the experiment at 8 K (representing $\sim$2\% of the depleted sulfur upon irradiation), and a $\sim$15\% higher at 50 K. 
Therefore, the vast majority ($\sim$95\%) of the depleted sulfur could not be tracked at the end of Experiments 10 and 11.

\begin{figure}
    \centering
    \includegraphics[width=0.525\linewidth]{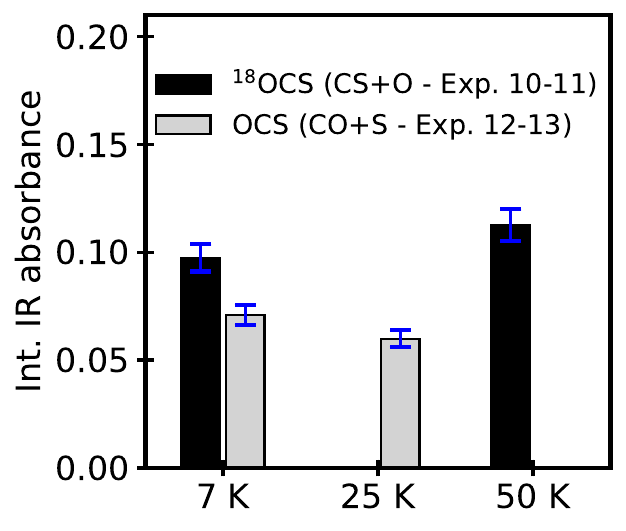}
    \caption{Formation of $^{18}$OCS (black) upon irradiation of H$_2^{18}$O:CS$_2$ ice samples, and OCS (gray) upon irradiation of CO:H$_2$S mixtures at 7$-$50 K, measured as the integration of the corresponding IR features.  
    The 5\% relative experimental uncertainty is indicated as blue error bars.
    }
    \label{fig:formation_ocs_h2o_h2s}
\end{figure}

Fig. \ref{fig:formation_ocs_h2o_h2s} shows the integrated $^{18}$OCS IR absorbances.  
As mentioned above, the $^{18}$OCS column density was $\sim$15\% higher at 50 K. 
A $\sim$15\% increase was also observed in the $^{18}$OCS formation through the CS+O pathway upon irradiation of a $^{13}$C$^{18}$O$_2$:CS$_2$ sample at 50 K, compared to irradiation at lower temperatures (Fig. \ref{fig:formation_ocs_co2_cs2}).

\subsection{2 keV electron irradiation of CO:H$_2$S ice samples at 6 and 25 K}\label{sec:co_h2s}

\begin{figure}
    \centering
    \includegraphics[width=1\linewidth]{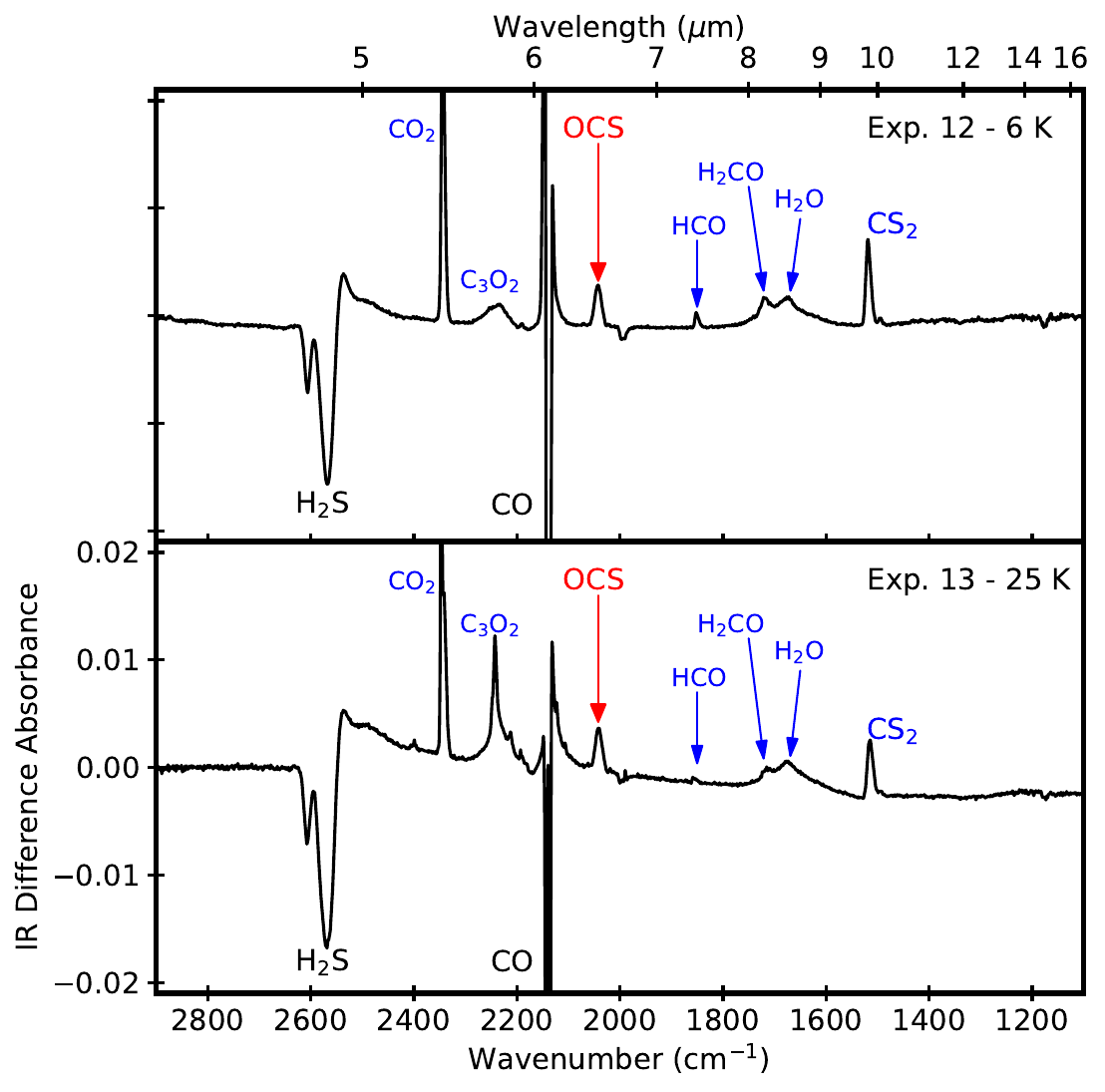}
    \caption{IR difference spectra obtained upon 2 keV electron irradiation of a CO:H$_2$S ice sample at 6 K (Exp. 12, top panel) and 25 K (Exp. 13, bottom panel).
    IR band assignments are indicated for the initial ice components (black), and ice chemistry products (blue), including OCS (red).}
    \label{fig:h2sco_ir}
\end{figure}


Following the results presented in Sect. \ref{sec:h2o_cs2}, we also studied the evolution with temperature of the OCS formation through the CO+S pathway in a different ice environment. 
To this purpose, Experiments 12 and 13 in Table \ref{tab:exp} consisted in the 2 keV electron irradiation of CO:H$_2$S ice samples at 6 and 25 K, respectively. 
In those experiments, OCS molecules probably formed through the CO + S atom-addition reaction (Eq. \ref{eq:co+s}), and/or through the CO + HS neutral-neutral reaction (Equations \ref{eq:co+hs}, \ref{eq:co+hs_2}, and \ref{eq:hsco+h}). 

As explained in Sect. \ref{sec:intro}, the energetic processing of CO:H$_2$S ice mixtures has been previously studied in \citet{ferrante08}, \citet{garozzo10}, and \citet{asper15}. 
A thorough analysis of the chemistry 
in Experiments 12 and 13 was thus beyond the scope of this paper. 
Our focus was instead in the dependence of OCS formation with temperature, that had not been evaluated in the above-mentioned works.   
Figure \ref{fig:h2sco_ir} shows the IR difference spectra in the 2900$-$1100 cm$^{-1}$ range for the experiments at 6 K (top panel) and 25 K (bottom panel), with OCS molecules detected at $\sim$2040 cm$^{-1}$. 
%
The integrated OCS IR absorbances are shown in Fig. \ref{fig:formation_ocs_h2o_h2s}. 
We observed a $\sim$15\% decrease in the OCS formation at 25 K compared to that at 6 K. 
%
Even though this decrease was lower than the $\sim$25\% decrease observed for the $^{13}$C$^{18}$O + S $\rightarrow$ $^{18}$O$^{13}$CS reaction in the $^{13}$C$^{18}$O$_2$:CS$_2$ experiments (Fig. \ref{fig:formation_ocs_co2_cs2}), and the $\sim$2.5 decrease observed in the $^{13}$C$^{18}$O:CS$_2$ experiments (Fig. \ref{fig:formation_ocs_co_cs2}), it was in line with the trend observed in those experiments.

\section{Discussion}\label{sec:disc}

\subsection{Relative contribution of the CO+S and CS+O pathways to the formation of OCS in interstellar ices}\label{sec:disc_ocs}

The experimental results presented in Sections \ref{sec:co2_cs2_7K}$-$\ref{sec:co_cs2_7K} suggest that the CS + $^{18}$O $\rightarrow$ $^{18}$OCS atom-addition reaction is more favorable than the $^{13}$C$^{18}$O + S $\rightarrow$ $^{18}$O$^{13}$CS reaction. 
In the irradiated $^{13}$C$^{18}$O$_2$:CS$_2$ ice samples 
(where both pathways were expected to equally contribute to the formation of OCS), 
up to $\sim$75\% of the detected OCS molecules were formed through the CS+O pathway 
(Sections \ref{sec:co2_cs2_7K} and \ref{sec:co2_cs2_vuv}). 
At the same time, in an irradiated $^{13}$C$^{18}$O:CS$_2$ ice sample (with a much higher abundance of $^{13}$C$^{18}$O compared to the other reactants), the relative contribution of both pathways was close to 50\% (Sect \ref{sec:co_cs2_7K}). 
%

According to the literature, photodissociation of 
CO$_2$ and CS$_2$ molecules with VUV photons mainly leads to 
CO(X$^1\Sigma^+_g$) + O($^1$D), 
and CS (a$^3\Pi$/A$^1\Pi$) + S($^3$P) \citep[respectively, see][]{slanger71,okabe78,zhu90}. 
As a result, in the VUV-photon irradiated $^{13}$C$^{18}$O$_2$:CS$_2$ ice samples  (Experiments 5 and 6 in Table \ref{tab:exp}), the reactants of the CS + $^{18}$O atom-addition reaction would be in electronic excited states (i.e., more reactive), while those of the  $^{13}$C$^{18}$O + S reaction  would be in the electronic ground state.   
%
On the other hand, \citet{zhu18} and \citet{maity13} reported that 5 keV electron irradiation of CO$_2$ and CS$_2$ ice molecules led to the formation of CO + O($^1$D/$^3$P), and CS(X$^1\Sigma^+_g$) + S($^3$P),  respectively. 
The same dissociation channels could be expected upon irradiation with 2 keV electrons 
(Experiments 1$-$4 in Table \ref{tab:exp}). 
Despite the different reported dissociation channels following VUV photon and keV electron irradiation of CO$_2$ and CS$_2$ molecules, the contribution of the CS+O and CO+S pathways to the formation of OCS molecules was very similar upon irradiation of $^{13}$C$^{18}$O$_2$:CS$_2$ ice samples with 2 keV electrons and VUV photons (Sections \ref{sec:co2_cs2_7K} and \ref{sec:co2_cs2_vuv}). 
This suggests that the electronic state of the reactants might not have played a significant role in the formation of OCS in the experiments reported in this work. 
%

Independently from the dissociation channels of $^{13}$C$^{18}$O$_2$, $^{13}$C$^{18}$O, and CS$_2$, 
theoretical calculations reported in \citet{adriaens10} indicated that oxidation of CS is more energetically favorable than sulfurization of CO. 
Even though \citet{adriaens10} studied the CO + S/HS and CS + O/OH reactions on a coronene surface instead that in the bulk of an ice mantle (see Sect. \ref{sec:intro}), some of their findings can be used to interpret the results of this work.  
Three types of surface reactions were studied in \citet{adriaens10}: 
the Langmuir-Hinshelwood mechanism (LH, where both reactants are in thermal equilibrium with the surface, i.e., in the vibrational ground state), 
the hot atom mechanism (HA, where at least one of the reactants is thermally activated), and 
the Eley-Rideal mechanism (ER, where one of the reactants is in the gas phase). 
While the latter is not relevant to ice bulk chemistry, the first two mechanisms could be comparable to the processes taking place in ice mantles. 
According to these theoretical calculations, the addition reaction of a S atom in the ground electronic state to CO (CO + S($^3$P)) can only take place to a significant extent in the ISM if one of the reactants is thermally activated (i.e., only the reaction through the HA mechanism is exothermic). 
%
In case the S atom is in the first electronic excited state, the CO + S($^1$D) reaction would be more exothermic than the CO + S($^3$P) reaction, and could in principle proceed through both, the LH and the HA mechanisms. 
However, S($^1$D) must also be thermally activated for the reaction to proceed, because otherwise it would chemisorb on the coronene surface, and the reaction would be strongly activated. 
On the other hand, the CS + O($^3$P/$^1$D) reactions were more exothermic than the equivalent CO + S ones. 
In addition, the calculated CS adsorption energy was more than two times higher than that of CO. This would reduce the efficiency of the CS desorption, leading to higher reaction rates. 
More importantly, the addition reaction of an O atom to CS led to the formation of a van der Waals complex that significantly reduced the activation barriers of these reactions, that could be easily overcome in the ISM \citep{adriaens10}.  
As a result, the CS + O($^3$P) reaction could take place in the ISM without thermal activation of the reactants, unlike the CO + S($^3$P) reaction. 
%
%
%
Therefore, previous theoretical calculations predicted that formation of OCS through the CS+O pathway would be more favorable than through the CO+S pathway. 
The results presented in this work thus represent the first experimental evidence that oxidation of CS would be preferred over sulfurization of CO in interstellar ice mantles.

\subsection{The effect of ice temperature on the CO+S and CS+O OCS formation pathways}\label{sec:disc_ocsT} 

The results presented in Sections \ref{sec:co2cs2_T} and \ref{sec:cocs2_T} indicated that, in $^{13}$C$^{18}$O$_2$:CS$_2$ and $^{13}$C$^{18}$O:CS$_2$ ice samples, 
the OCS formation through the $^{13}$C$^{18}$O + S reaction gradually took place to a lower extent as the ice temperature increased in the 7$-$50 K range. 
At the same time, 
formation through the CS + $^{18}$O was slightly higher at 50 K. 
Similar trends were observed in  
H$_2^{18}$O:CS$_2$ and CO:H$_2$S ice samples, where the neutral-radical reactions CS + $^{18}$OH and CO + HS could also contribute to the formation of OCS (Sections \ref{sec:h2o_cs2} and \ref{sec:co_h2s}, respectively). 
%

As mentioned in Sections \ref{sec:co2cs2_T} and \ref{sec:cocs2_T}, the gradual decrease observed 
for the $^{13}$C$^{18}$O + S reaction 
was probably not related to a lower availability of $^{13}$C$^{18}$O molecules in the ice.
In the $^{13}$C$^{18}$O$_2$:CS$_2$ experiments, this decrease was observed already at 25 K, with the same measured $^{13}$C$^{18}$O ice column density as at 7 K 
(Sect. \ref{sec:co2cs2_T}). 
In the $^{13}$C$^{18}$O:CS$_2$ experiments, the formation of other products derived from $^{13}$C$^{18}$O, such as $^{13}$C$^{18}$O$_2$ and $^{13}$C$_3^{18}$O$_2$, remained constant at all temperatures (Sect. \ref{sec:cocs2_T}). 
We speculate that this decrease could instead be due to a lower availability of S atoms at higher temperatures.  
This could be caused, for example, by a subtle increase in the efficiency of sulfur allotrope formation (see Sect. \ref{sec:disc_chem}). 
Unfortunately, we could not confirm this scenario with our experimental results because the observed decrease, while significant in terms of OCS formation, represented a very small change in the global balance of the sulfur chemistry (due to the small conversion from CS$_2$ to OCS in these experiments). 
%
%
This possibility will be further explored in a follow-up paper using theoretical simulations to model the irradiation experiments presented in this work \citep{olli24}. 

\subsection{Sulfur chemistry in CS$_2$-bearing ices}\label{sec:disc_chem}

\begin{figure}
    \centering
    \includegraphics[width=0.7\linewidth]{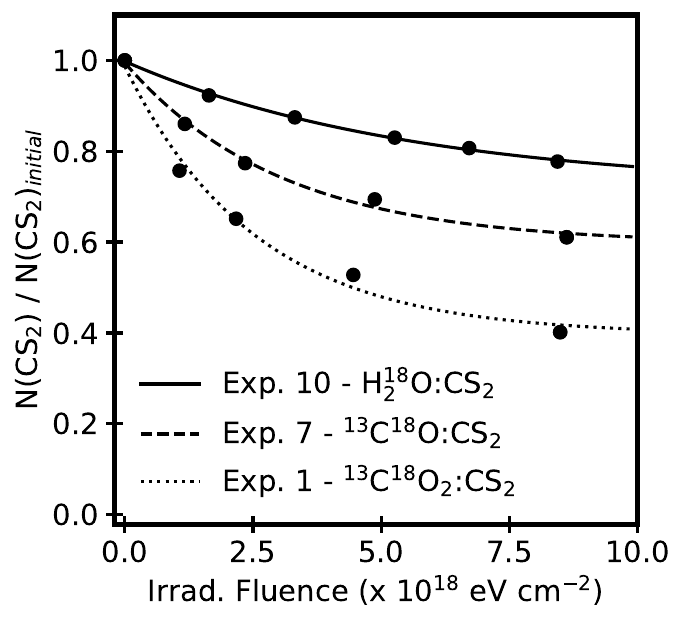}
    \caption{Evolution of the CS$_2$ column density (normalized to the initial CS$_2$ column density, black circles) with the irradiated fluence in H$_2^{18}$O:CS$_2$ (Exp. 10, solid line), $^{13}$C$^{18}$O:CS$_2$ (Exp. 7, dashed line), and $^{13}$C$^{18}$O$_2$:CS$_2$ (Exp. 1, dotted line) ice samples. 
    Black lines correspond to pseudo-first order fits.}
    \label{fig:cs2_destruction}
    
\end{figure}
\begin{figure*}
    \centering
    \includegraphics[width=0.305\linewidth]{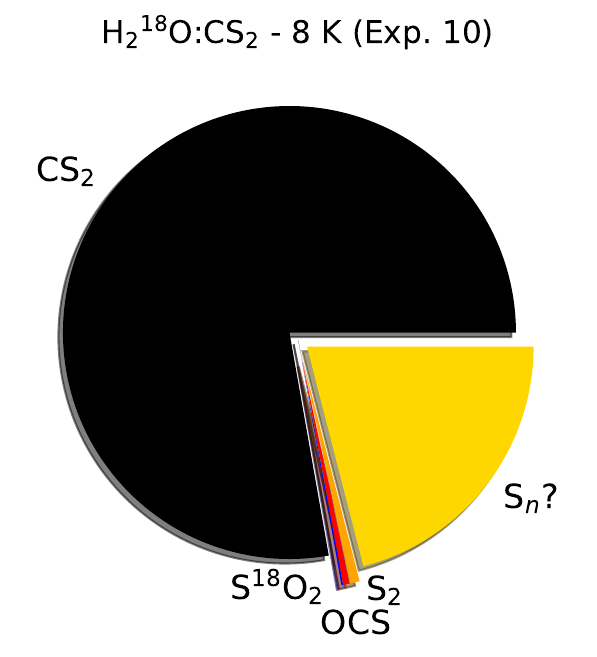}
    \includegraphics[width=0.3\linewidth]{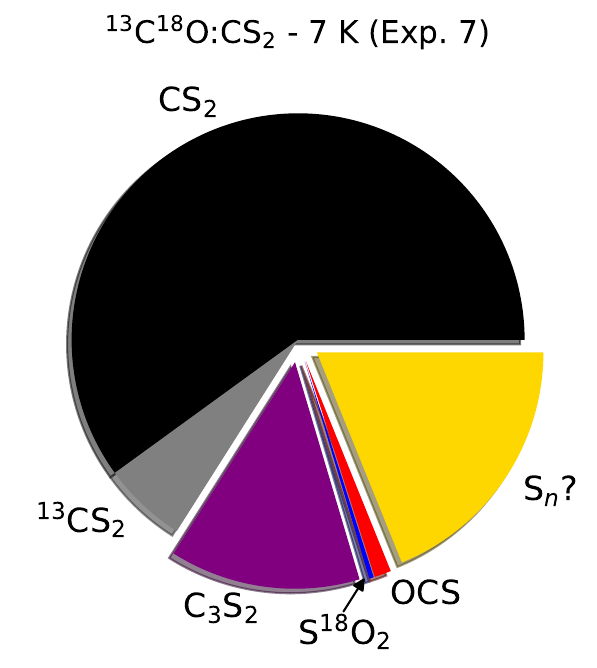}
    \includegraphics[width=0.335\linewidth]{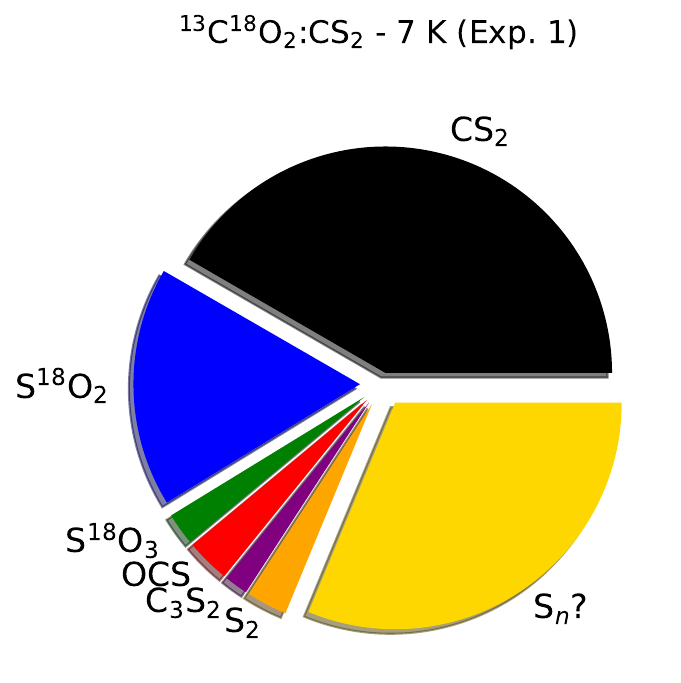}
    \caption{Distribution of the initial sulfur in the ice across the remaining CS$_2$ molecules and the detected S-bearing products after 2 keV electron irradiation of H$_2^{18}$O:CS$_2$ (Exp. 10, left panel), $^{13}$C$^{18}$O:CS$_2$ (Exp. 7, middle panel), and $^{13}$C$^{18}$O$_2$:CS$_2$ (Exp. 1, right panel) ice samples at 7$-$8K.
    The OCS and C$_3$S$_2$ slabs include all detected isotopologs of each species.}
    \label{fig:s_chemistry}
\end{figure*}

As explained in Sect. \ref{sec:intro}, Experiments 1$-$11 in Table \ref{tab:exp} also allowed us to explore the sulfur chemistry of ices containing CS$_2$ molecules in realistic environments, namely, H$_2$O-, CO-, and CO$_2$-rich ice matrices. 
Fig. \ref{fig:cs2_destruction} shows the evolution of the CS$_2$ ice column density with respect to the initial value during 2 keV electron irradiation of H$_2^{18}$O:CS$_2$, $^{13}$C$^{18}$O:CS$_2$, and $^{13}$C$^{18}$O$_2$:CS$_2$ ice samples at 7$-$8 K (corresponding to Experiments 10, 7, and 1 in Table \ref{tab:exp}, respectively). 
In all experiments, the depletion of CS$_2$ after irradiation was close to the steady state value. 
Dissociation of CS$_2$ molecules led to the formation of several S-bearing products. 
Figure \ref{fig:s_chemistry} shows the distribution of S atoms after irradiation in these experiments, including the remaining CS$_2$ molecules and the different S-bearing products. 
Figures \ref{fig:cs2_destruction} and \ref{fig:s_chemistry} indicate that sulfur chemistry proceeded to a higher extent in the $^{13}$C$^{18}$O$_2$:CS$_2$ ice samples, where $\sim$60\% of the initial CS$_2$ molecules were transformed into other products, 
compared to $\sim$40\% in $^{13}$C$^{18}$O:CS$_2$ ices, and $\sim$20\% in H$_2^{18}$O:CS$_2$ samples. 

The most abundant S-bearing product detected in the irradiated $^{13}$C$^{18}$O$_2$:CS$_2$ ice sample was S$^{18}$O$_2$, that accounted for $\sim$26\% of the sulfur participating in the ice chemistry (i.e., not present in CS$_2$ molecules after irradiation), while OCS molecules represented $\sim$5\% of this fraction of sulfur. 
Interestingly, $\sim$60\% of the sulfur atoms involved in the chemistry could not be detected at the end of the experiment (see right panel of Fig. \ref{fig:s_chemistry}). 
A similar problem was encountered in \citet{asper15}, with detected S-bearing products accounting for less than 20\% of the initial sulfur content after photolysis of CO:H$_2$S and CO$_2$:H$_2$S ices. 
Similarly, \citet{mifsud22} reported that 25$-$45\% of the initial sulfur was not observable after electron irradiation of amorphous and crystalline H$_2$S and SO$_2$ ice samples.
As in \citet{asper15} and \citet{mifsud22}, we speculate that the missing sulfur in our experiments could be contained in long sulfur allotropes (S$_n$) undetectable with our instruments.
Formation of S-polymers from S-bearing species was first suggested in \citet{barnes74}. 
Unfortunately, these molecules are not usually active in the IR, and their detection is elusive. 
\citet{guillermo02} was able to detect S-polymers in the room-temperature residues resulting from the UV irradiation of H$_2$S ice samples using chromatographic techniques. 
Formation of (semi-)refractory sulfur chains in H$_2$S-bearing ices, including sulfur allotropes and polysulfides (H$_2$S$_x$), has more recently been studied in \citet{cazaux22} and \citet{hector24}. 
In particular, \citet{hector24} reported the tentative detection of the sulfur allotropes S$_2$ and S$_8$ during the TPD of an irradiated H$_2$O:H$_2$S ice at 10 K. 
%
In our experiments, thermal desorption of S$_2$ was detected at 110$-$115 K (Sect. \ref{sec:co2_cs2_7K}). 
Unfortunately, the molecular mass of sulfur allotropes with 4 or more atoms fell outside of the QMS mass range. 
In any case, S$_8$ could have been detected through the S$_2^+$ fragment, but the desorption temperature reported in \citet{hector24} ($>$260 K) was beyond the temperature range of our TPD. 

The fraction of missing sulfur in our experiments was somewhat lower ($\sim$45\%) in the irradiated $^{13}$C$^{18}$O:CS$_2$ ice  
(where thermal desorption of S$_2$ was not detected, Sect. \ref{sec:co_cs2_7K}). 
In this case, C$_3$S$_2$ and its isotopologs were the most abundant detected S-bearing product, accounting for $\sim$34\% of the sulfur contained in products, while OCS molecules represented $\sim$3\% of this sulfur. 
%
On the other hand, OCS was the most abundant S-bearing product detected in irradiated H$_2^{18}$O:CS$_2$ samples, representing 
$\sim$2\% of the sulfur participating in the chemistry. 
However, most of this sulfur 
($\sim$95\%) could not be detected after irradiation (left panel of Fig. \ref{fig:s_chemistry}). 
As in the $^{13}$C$^{18}$O$_2$:CS$_2$ experiments, we speculate that the missing sulfur in these experiments could be contained in long, undetectable sulfur chains. 
%
%
We note that, in the literature, irradiation of H$_2$O:H$_2$S ice samples mainly led to the formation of polysulfides (H$_2$S$_x$) and sulfur allotropes, suggesting that reactions between S-bearing reactants could have dominated the chemistry over reactions with H$_2$O molecules and OH radicals \citep{antonio11,hector24}. 
A similar scenario could have occurred in the H$_2^{18}$O:CS$_2$ experiments presented in this work. 

\subsection{Astrophysical implications}\label{sec:disc_astimp}


%
The experimental results presented in this work 
indicate that, in irradiated $^{13}$C$^{18}$O$_2$:CS$_2$ ice samples at 7$-$8 K, the contribution of the CS+O pathway to the formation of OCS was $\sim$3 times higher than the contribution of the CO+S pathway. 
This value increased up to a factor of $\sim$6 at 50 K. 
In interstellar ice mantles, the relative contribution of the two pathways will depend not only on the relative efficiency of the reactions, but also on the availability of the reactants and, in particular, on the CO/CS and O/S abundance ratios. 
In our experiments, dissociation of the $^{13}$C$^{18}$O$_2$ and CS$_2$ molecules led to the same $^{13}$C$^{18}$O/CS and atomic $^{18}$O/S ratios. 
In the ISM, the gas-phase CO/CS ratio is $\sim$2 $\times$ 10$^4$ \citep[and references therein]{laas19}, 
which would favor formation of OCS through the CO+S pathway if the same ratio was found in interstellar ices. 
Even though CS is yet to be detected in interstellar ices, it has been observed in comets (whose composition is expected to resemble that of interstellar ices), with much lower CO/CS ratios ($\sim$10$-$100) than those measured in the gas phase \citep[][]{bockelee17}. 
In addition, the estimated CS binding energy is higher than that of CO \citep[see, e.g.,][]{wakelam17}. 
As a result, ice mantles in warm interstellar environments 
could present even lower CO/CS ratios.
Under these circumstances, the more favorable CS+O pathway could play a significant role in the formation of OCS. 
On the other hand, the volatile O/S ratio in the solid phase is not fully constrained. 
If a large fraction of S is locked in (semi-)refractory form \citep[as discussed in][]{millar90,ruffle99,vidal17,drozdovskaia18,kama19,fuente19,pablo20,fuente23}, 
that could also favor a larger contribution of the CS+O pathway. 
%
%


In an attempt to shed light on the molecular environment and formation history of OCS in interstellar ices, \citet{boogert22} recently evaluated the observed ice column density correlations between OCS and different species. 
However, the results were not conclusive. 
The best correlations were found with OCN$^-$ and CH$_3$OH. 
This would suggest that formation of OCS takes place in the apolar, CO-rich ice layer, thus probably through a CO+S pathway. 
However, no correlation was found between OCS and the apolar component of CO.
Moreover, even though a somewhat poorer correlation was found with the polar component of CO (that would be present in the H$_2$O-rich ice layer), the OCS ice column density did not correlate with H$_2$O either. 
%
In our experiments, OCS formation was readily observed in a H$_2$O-rich, a CO-rich, and a CO$_2$-rich ice environment, with conversions on the same order of magnitude in all cases (representing $\sim$2$-$5\% of the sulfur coming from dissociated CS$_2$ molecules, Sect. \ref{sec:disc_chem}). 
In contrast, other detected S-bearing products presented differences in their conversions of up to an order of magnitude across different ice environments. 
Following the results presented in this work, formation of OCS has now been observed in the laboratory upon energetic processing of virtually any ice mixture containing O, C, and S \citep[CO:H$_2$S, CO$_2$:H$_2$S, CO:SO$_2$, CO$_2$:SO$_2$, CO:CS$_2$, CO$_2$:CS$_2$, H$_2$O:CS$_2$, O$_2$:CS$_2$, see][and this work]{ferrante08,garozzo10,maity13,asper15}. 
The robust formation of this molecule 
in different ice environments may explain its ubiquitous detection in interstellar ice mantles \citep[see, e.g.,][]{boogert22}. 

Finally, the experiments presented in this work suggest that energetic processing of CS$_2$-bearing ices might lead to the formation of sulfur allotropes. 
These molecules are one of the potential carriers proposed in the literature to harbor the missing sulfur in dense interstellar regions \citep[see,e.g.,][]{wakelam04,fuente19,shingledecker20,cazaux22}. 
\citet{fuente19} reported that depletion of sulfur by one order of magnitude is already observed in translucent clouds, where sulfur atoms or ions would be incorporated onto the surface of dust grains.  
According to the theoretical model presented in \citet{cazaux22}, this depleted sulfur would preferentially form long S-chains. 
Sulfur is further depleted in the interior of dense clouds, where formation of S-polymers could also take place upon energetic processing of S-bearing ice mantles. 
Thus far, formation of (semi-)refractory sulfur chains had been observed 
in irradiated ices containing H$_2$S \citep[see, e.g.,][]{guillermo02,garozzo10,antonio11,asper15,cazaux22,hector24} 
and, likely, SO$_2$ \citep{mifsud22}. 
Therefore, the results presented in this work could represent 
an additional evidence that formation of sulfur allotropes is common in energetically processed S-bearing ice mantles.
Further experiments beyond the scope of this paper are needed in order to confirm and characterize the formation of sulfur allotropes in CS$_2$-bearing ices.

\section{Conclusions}\label{sec:conc}

In this work, we have simulated the energetic processing of $^{13}$C$^{18}$O$_2$-, $^{13}$C$^{18}$O-, and H$_2^{18}$O-rich, CS$_2$-bearing ice samples at temperatures relevant to the dense ISM (7$-$50 K). 

\begin{itemize}

\item Sulfur chemistry proceeded to a larger extent in $^{13}$C$^{18}$O$_2$:CS$_2$ ices, where depletion of the initial CS$_2$ molecules was up to 3 times higher than in $^{13}$C$^{18}$O:CS$_2$ and H$_2^{18}$O:CS$_2$ samples.

\item Dissociation of CS$_2$ molecules led to the formation of a variety of S-bearing products, including S$^{18}$O$_2$, C$_3$S$_2$, OCS, and S$_2$. Formation of OCS has been readily observed in the laboratory upon energetic processing of ice samples with different S-carriers and compositions, which may explain its ubiquitous detection in interstellar ices. 

\item In $^{13}$C$^{18}$O$_2$:CS$_2$ ice samples irradiated at 7 K (with the same $^{13}$C$^{18}$O/CS and atomic $^{18}$O/S ratios), the CS+O pathway was responsible for the formation of $\sim$75\% of the OCS molecules, while the CO+S pathway only contributed to $\sim$25\% of the total OCS formation. 
At the same time, in irradiated $^{13}$C$^{18}$O:CS$_2$ ice samples (with a much higher abundance of $^{13}$C$^{18}$O compared to the other reactants), the relative contribution of both pathways was close to 50\%.
The measured relative contribution did not depend on the irradiation source (2 keV electrons or VUV photons). 

\item In addition, the contribution of the CS+O pathway increased at higher temperatures, due to a combination of a slight increase in the formation of OCS through the CS+O pathway at $\sim$50 K, and a gradual decrease of the CO+S pathway in the 7$-$50 K temperature range.
%
%
The latter could be due to a lower availability of S atoms to participate in the formation of OCS at higher temperatures, perhaps due to a subtle increase in the efficiency of competing reactions, such as the formation of sulfur allotropes. 

\item These experimental results thus suggested that the CS+O pathway was more favorable (as theoretically predicted in the literature), and could play a role in OCS formation in interstellar ices, especially in warm regions where CO would preferentially be in the gas phase. 

\item More than 50\% of the sulfur atoms involved in the ice chemistry were not detected at the end of the experiments, and could be contained in long, undetectable sulfur allotropes. These molecules are one of the potential S-carriers in the dense ISM. 
These results could represent 
an additional evidence that formation of sulfur allotropes is common in energetically processed S-bearing ice mantles.
Further experiments are needed in order to confirm and characterize the formation of sulfur allotropes in CS$_2$-bearing ices. 

\end{itemize}


\section*{Acknowledgements}
The project leading to these results has received funding from “la Caixa” Foundation, under agreement LCF/BQ/PI22/11910030. 
This work was also supported by a grant from the Simons Foundation (686302, K\"O) and an award from the Simons Foundation (321183FY19, K\"O). 
GMMC and HC received funding from project PID2020-118974GB-C21 by the Spanish Ministry of Science and Innovation.
AF has received funding from the European Research Council (ERC) under the European Union’s Horizon Europe research and innovation programme ERC-AdG-2022 (GA No. 101096293)
AF also thanks project PID2022-137980NB-I00 funded by the Spanish Ministry of Science and Innovation/State Agency of Research MCIN/AEI/ 10.13039/501100011033 and by “ERDF A way of making Europe”.

\section*{Data Availability}
The data underlying this article are available in the \texttt{zenodo} repository at https://zenodo.org, and can be accessed with DOI: 10.5281/zenodo.13908992. 
 







\appendix




\section{Additional plots of the 2 keV electron irradiation of $^{13}$C$^{18}$O$_2$:CS$_2$ ice samples at 7$-$50 K}\label{app:co2_cs2}

\begin{figure}
    \centering
    \includegraphics[width=1\linewidth]{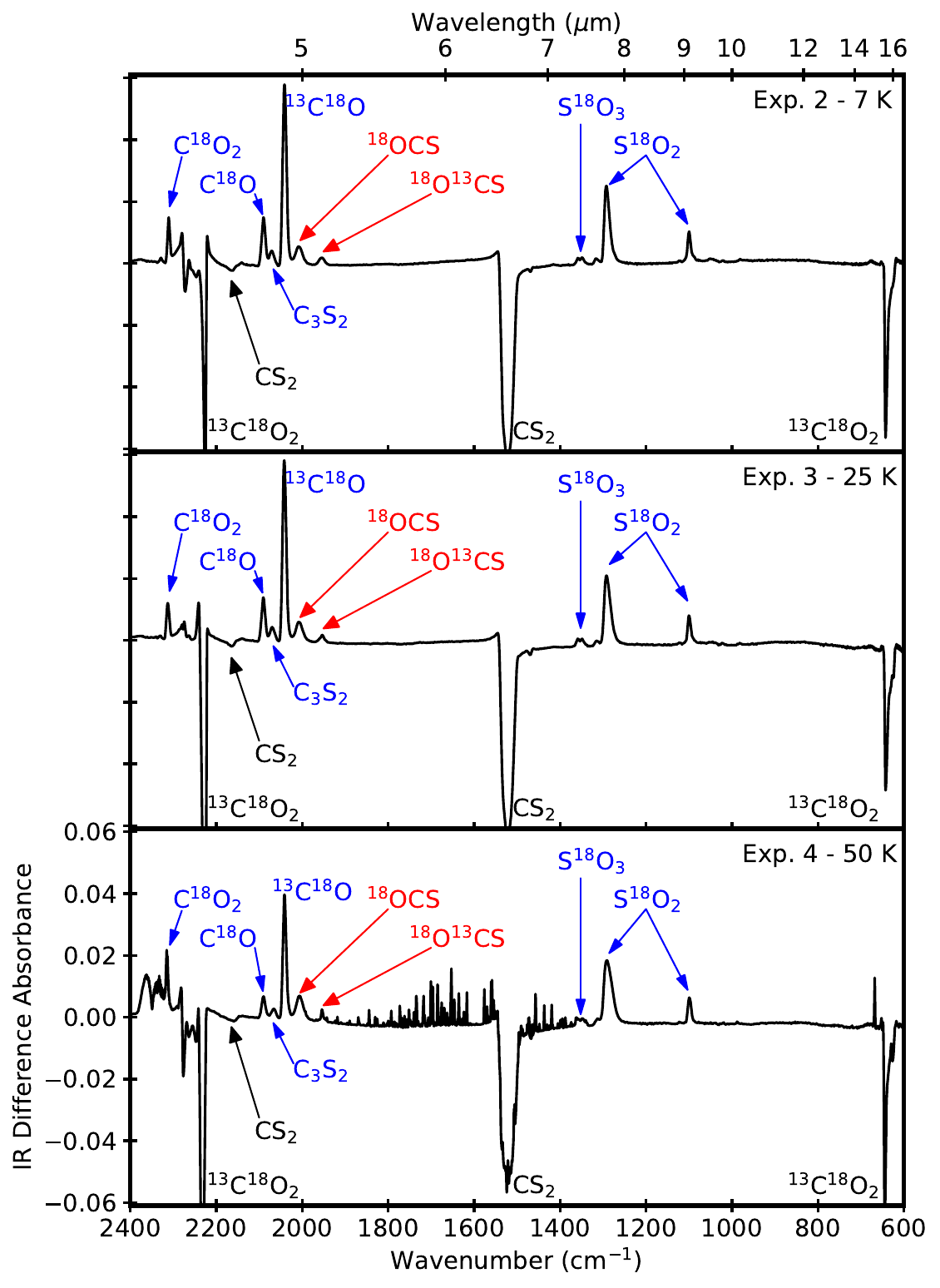}
    \caption{IR difference spectrum obtained upon 2 keV electron irradiation of a $^{13}$C$^{18}$O$_2$:CS$_2$ ice sample at 7 K (Exp. 2, top panel), 25 K (Exp. 3, middle panel), and 50 K (Exp. 4, bottom panel). 
    Narrow features in the bottom panel are due atmospheric contamination in the IR detector chamber in this particular experiment. 
    IR band assignments are indicated for the initial ice components (black), and ice chemistry products (blue), including the OCS isotopologs (red).}
    \label{fig:co2_cs2_IR_allT}
\end{figure}

The IR difference spectra in the 2400$-$600 cm$^{-1}$ range obtained upon 2 keV electron irradiation of $^{13}$C$^{18}$O$_2$:CS$_2$ ice samples at 7 K (Exp. 2), 25 K (Exp. 3), and 50 K (Exp. 4) are shown in Figure \ref{fig:co2_cs2_IR_allT}. 
The spectra did not present significant differences in this temperature range. 

\begin{figure*}
    \centering
    \includegraphics[width=0.6\linewidth]{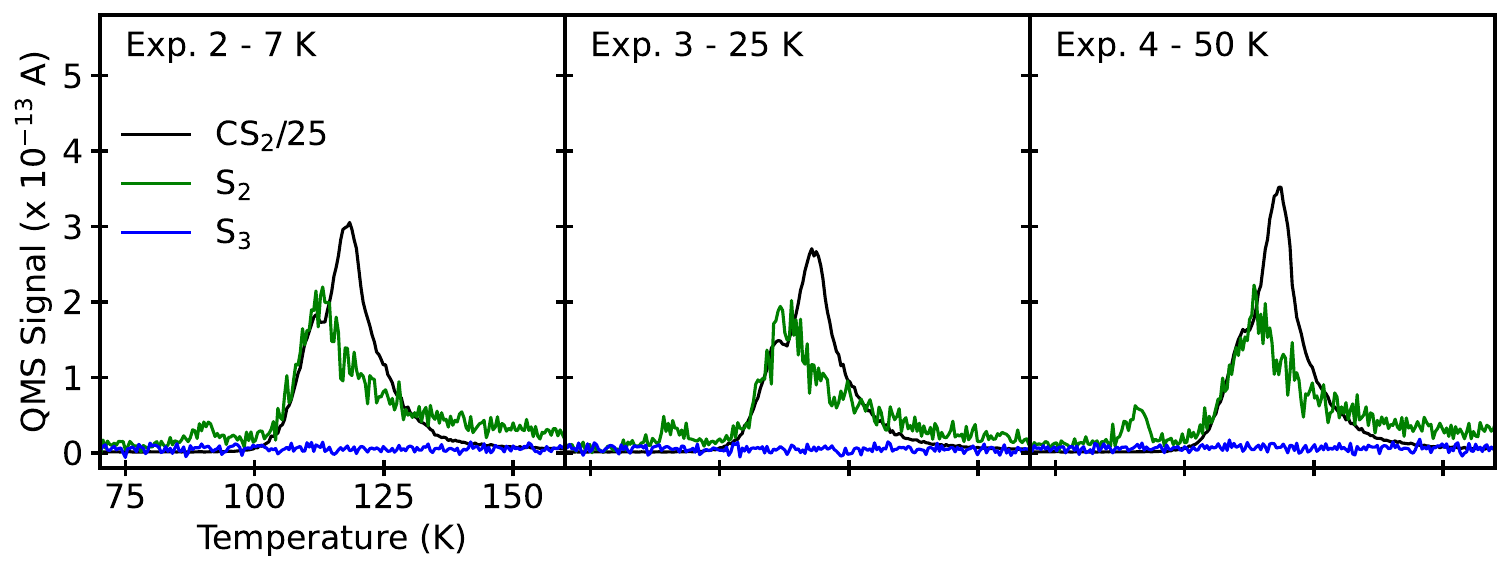}
    \includegraphics[width=0.6\linewidth]{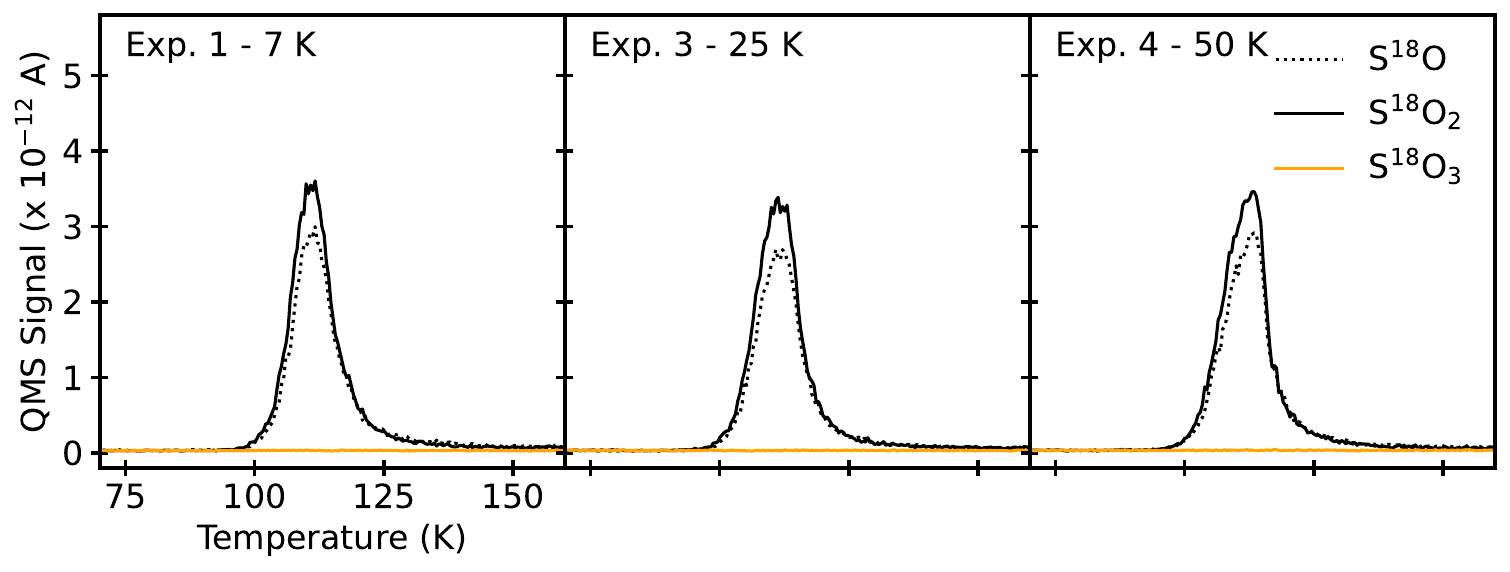}
    \caption{Top: TPD curves corresponding to CS$_2$ (black), S$_2$ (green) and S$_3$ (blue) molecules measured during thermal desorption of the $^{13}$C$^{18}$O$_2$:CS$_2$ ice samples irradiated at 7 K (Exp. 1, left panel), 25 K (Exp. 3, middle panel) and 50 K (Exp. 4, right panel). Bottom: TPD curves corresponding to S$^{18}$O$_2$ (solid black), S$^{18}$O$_3$ (orange), and the S$^{18}$O fragment (dotted black) during thermal desorption in Experiments 2 (irradiated at 7 K, left panel), 3 (25 K, middle panel), and 4 (50 K, right panel).}
    \label{fig:co2cs2_tpd_so2}
    \label{fig:tpd_s2_allT}
\end{figure*}

In addition to the thermal desorption of $^{18}$OCS and $^{18}$O$^{13}$CS shown in the main body of the manuscript (Fig. \ref{fig:co2cs2_tpd_ocs}, bottom panels), desorption of the S-bearing products SO$_2$ and S$_2$ was also observed during warm-up of the irradiated ices. The corresponding TPD curves for the ice samples irradiated at 7 K (Exp. 1), 25 K (Exp. 3), and 50 K (Exp. 4) are shown in the top and bottom panels of Fig. \ref{fig:co2cs2_tpd_so2}.  
The small fraction of S$^{18}$O$_3$ molecules produced upon irradiation (Sect. \ref{sec:co2_cs2_7K}) probably fragmented in the QMS and contributed to the S$^{18}$O$_2$ and S$^{18}$O signals. Thermal desorption of S$^{18}$O molecules from the ice could not be confirmed with the available data.
Thermal desorption of C$_3$S$_2$ was not detected during warm-up of the irradiated $^{13}$C$^{18}$O$_2$:CS$_2$ ice samples, but it was observed in the $^{13}$C$^{18}$O:CS$_2$ experiments (Fig. \ref{fig:cocs2_tpd_c3s2}, bottom panels). 

\section{Desorption of molecules during irradiation of $^{13}$C$^{18}$O$_2$:CS$_2$ ice samples}\label{app:kco}

As explained in Sect. \ref{sec:co2cs2_T}, desorption of $^{13}$C$^{18}$O, C$^{18}$O molecules and $^{13}$C$^{18}$O$_2$ molecules was observed during irradiation of the $^{13}$C$^{18}$O$_2$:CS$_2$ ice samples (Fig. \ref{fig:co2cs2_irr_qms}). 
This desorption was selective (no other species desorbed during irradiation), and took place only when the ice was being actively irradiated (the corresponding QMS signal decreased down to the background level during the irradiation pauses, as it occurs in VUV photodesorption experiments). 
The desorption of $^{13}$C$^{18}$O and C$^{18}$O increased with the irradiation time in all experiments (probably due to the accumulation of molecules as they were being formed in the ice), and was more than an order of magnitude higher in the experiment at 50 K.

The QMS calibration process described in \citet{martin24} should in principle be used only with integrated TPD curves, since the $k_{CO}$ proportionality constant was derived from TPD calibration experiments. 
The desorption of molecules during irradiation of the $^{13}$C$^{18}$O$_2$:CS$_2$ ice samples might have followed different dynamics than the thermal desorption during warm-up of the ice samples, resulting in a different $k_{CO}$ value. We note that, for example, $k_{CO}$ derived from TPD experiments is different for different heating rates.  
In any case, we used the $k_{CO}$ value presented in Sect. \ref{sec:exp_TPD} to estimate the column density of $^{13}$C$^{18}$O molecules desorbing during irradiation in Experiments 1$-$4 as a first approximation. 
To this purpose, we divided $k_{CO}$ by the 2 K min$^{-1}$ heating rate applied in the calibration experiments to obtain a new value of 5.75 $\times$ 10$^{-12}$ A min ML$^{-1}$, 
leading to the results presented in Sect. \ref{sec:co2cs2_T}. 

\section{Additional plots of the 2 keV electron irradiation of $^{13}$C$^{18}$O:CS$_2$ ice samples at 7$-$25 K}\label{app:co_cs2}

\begin{figure}
    \centering
    \includegraphics[width=1\linewidth]{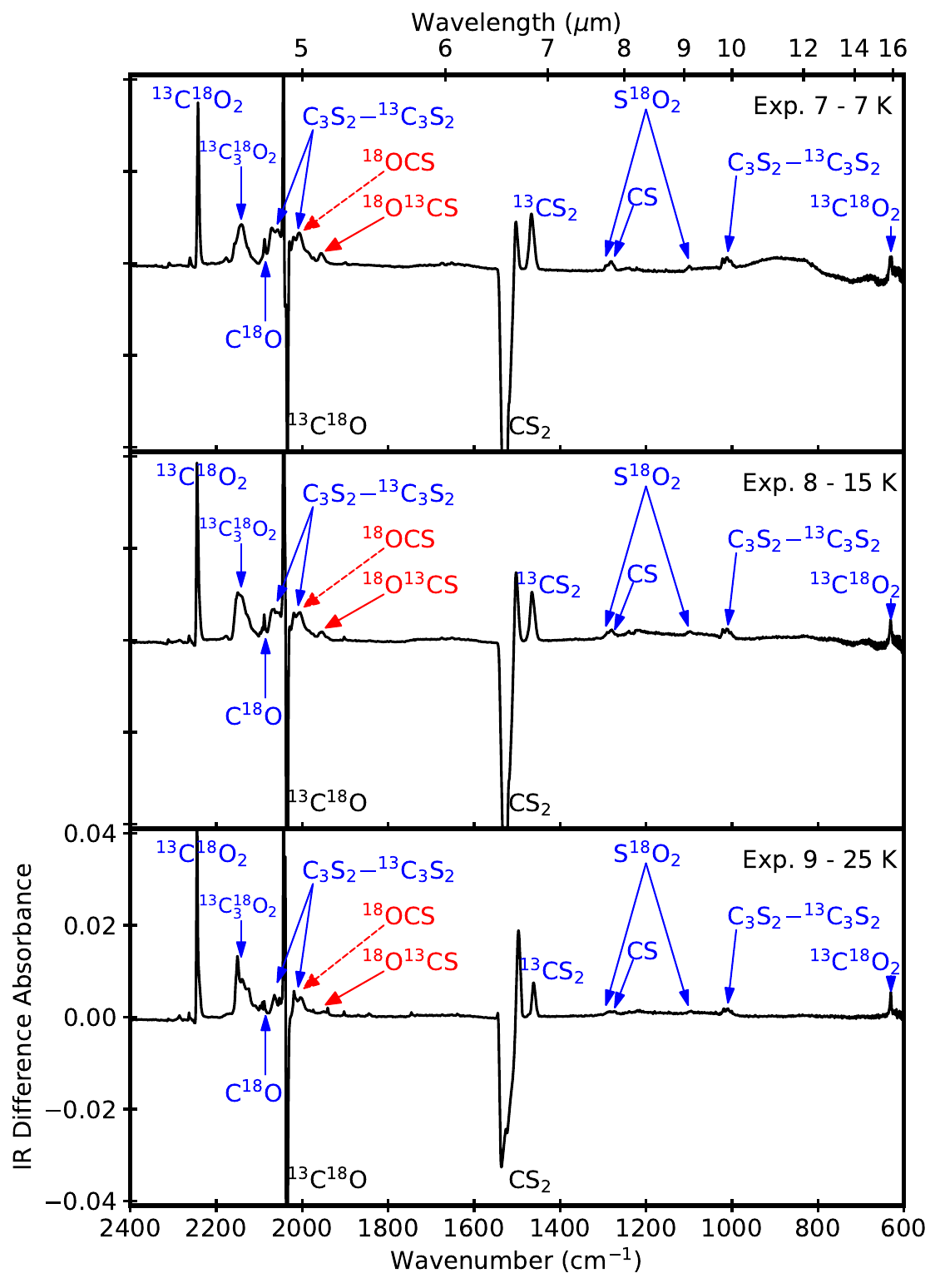}
    \caption{IR difference spectra obtained upon 2 keV electron irradiation of a $^{13}$C$^{18}$O:CS$_2$ ice sample at 15 K (Exp. 8, middle panel) and 25 K (Exp. 9, 25 K).  
    The IR difference spectrum of the ice sample irradiated at 7 K (Exp. 7, top panel, also shown in Fig. \ref{fig:cocs2_IR}) is presented for comparison. 
    IR band assignments are indicated for the initial ice components (black), and ice chemistry products (blue), including the OCS isotopologs (red).}
    \label{fig:cocs2_IR_allT}
\end{figure}

The IR difference spectra in the 2400$-$600 cm$^{-1}$ range obtained upon 2 keV electron irradiation of $^{13}$C$^{18}$O$_2$:CS$_2$ ice samples at 7 K (Exp. 7), 15 K (Exp. 8), and 25 K (Exp. 5) are shown in Figure \ref{fig:cocs2_IR_allT}. 
The main differences between these spectra are shown in the top panels of Fig. \ref{fig:cocs2_gauss}, and discussed in the main body of the manuscript. 

\begin{figure*}
    \centering
    \includegraphics[width=0.6\linewidth]{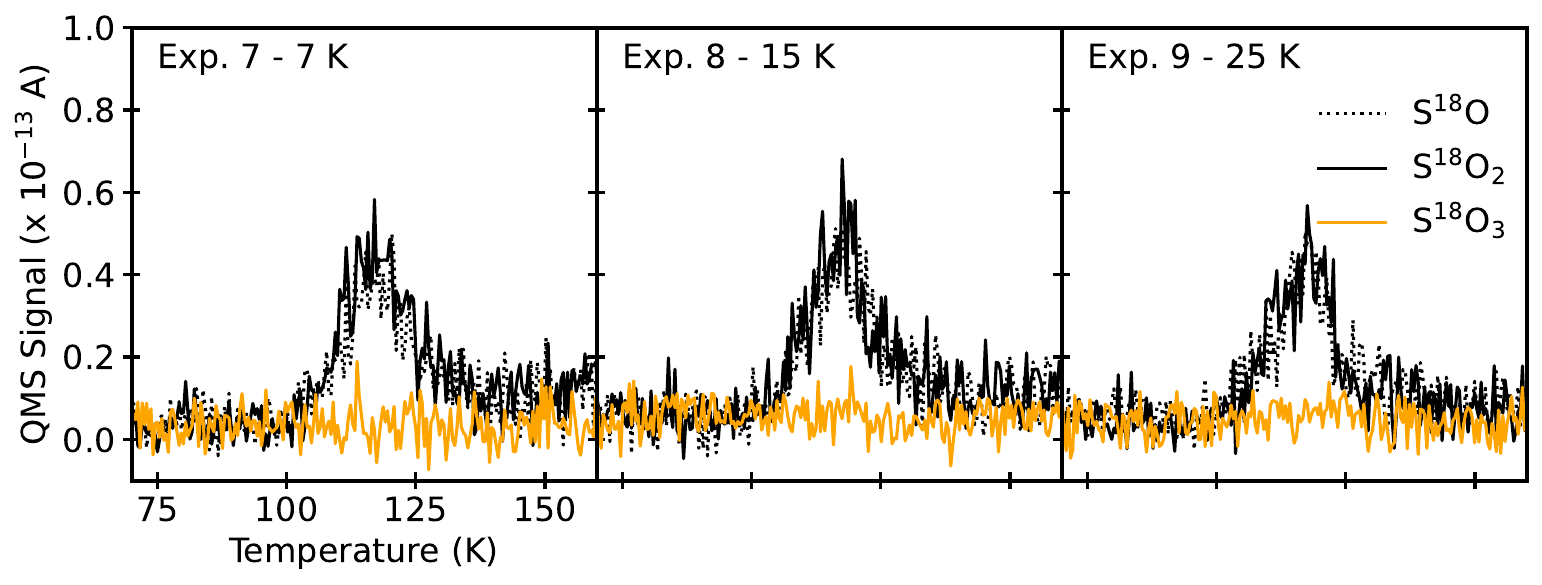}
    \includegraphics[width=0.6\linewidth]{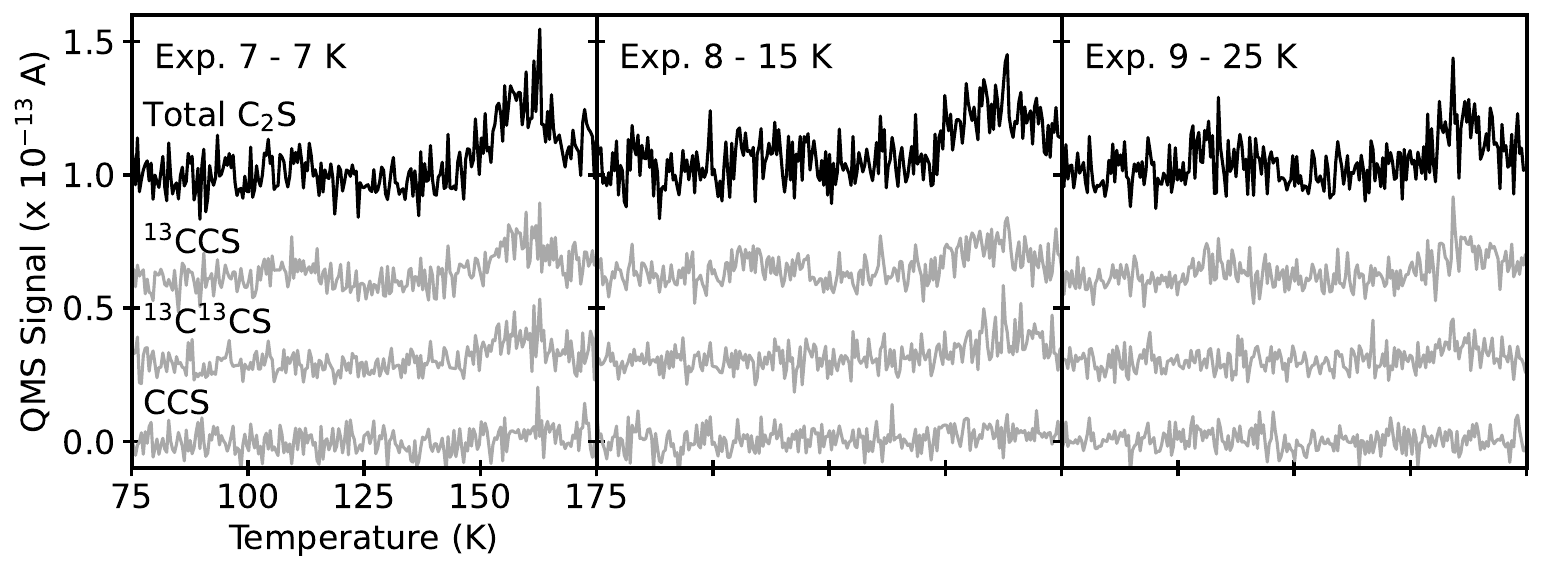}
    \caption{Top: TPD curves corresponding to S$^{18}$O$_2$ (solid black), S$^{18}$O$_3$ (orange), and the S$^{18}$O fragment (dotted black) measured during thermal desorption of the $^{13}$C$^{18}$O:CS$_2$ ice samples irradiated at 7 K (Exp. 7, left panel), 15 K (Exp. 8, middle panel) and 25 K (Exp. 9, right panel).
    Bottom: TPD curves corresponding to C$_2$S fragments of C$_3$S$_2$ isotopologs with different isotopic composition measured in the same experiments.The TPD curves are shifted for clarity.
    Note that for similar abundances of the different C$_3$S$_2$ isotopologs (Sect. \ref{sec:co_cs2_7K}), the C$_2$S fragment with one $^{13}$C was more abundant for statistical reasons.}
    \label{fig:cocs2_tpd_so2}
    \label{fig:cocs2_tpd_c3s2}
\end{figure*}

In addition to the thermal desorption of $^{18}$OCS and $^{18}$O$^{13}$CS shown in the bottom panels of Fig. \ref{fig:cocs2_tpd_ocs}, desorption of the S-bearing products SO$_2$ and C$_3$S$_2$ was also observed during warm-up of the irradiated ices. The corresponding TPD curves for the ice samples irradiated at 7 K (Exp. 7), 15 K (Exp. 8), and 25 K (Exp. 9) are shown in Fig. \ref{fig:cocs2_tpd_so2}. 

\section{Additional plots of the 2 keV electron irradiation of H$_2^{18}$O:CS$_2$ ice samples at 8$-$50 K}\label{app:h2o_cs2}

\begin{figure}
    \centering
    \includegraphics[width=1\linewidth]{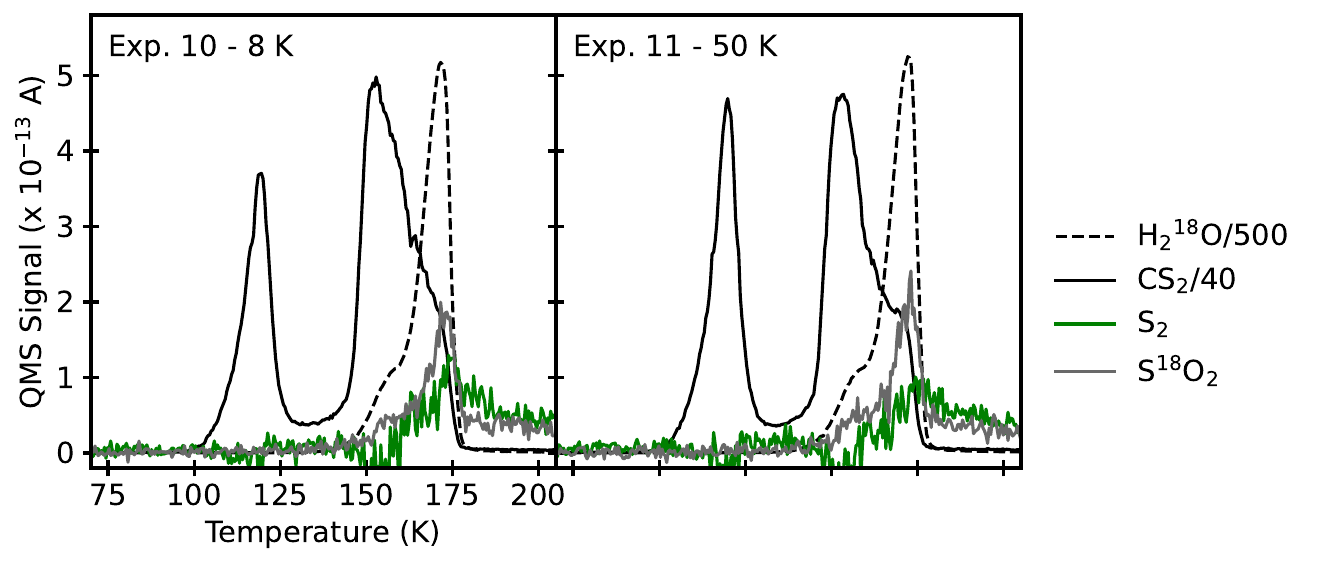}
    \caption{TPD curves corresponding to the remaining H$_2^{18}$O (dashed black) and CS$_2$ (solid black), and the produced S$^{18}$O$_2$ (gray) and S$_2$ (green), measured during thermal desorption of the H$_2^{18}$O:CS$_2$ ice samples irradiated at 8 K (Exp. 10, left panel) and 50 K in (Exp. 11, right panel). 
    }
    \label{fig:tpd_h2o_cs2_app}
\end{figure}

In addition to the thermal desorption of $^{18}$OCS (Fig. \ref{fig:tpd_h2o_cs2}), we also observed thermal desorption of S$^{18}$O$_2$ and S$_2$ during the TPD of the  H$_2^{18}$O:CS$_2$ ice samples irradiated at 8 and 50 K. 
The corresponding TPD curves are presented in Fig. \ref{fig:tpd_h2o_cs2_app}. 

\bsp	
\label{lastpage}
\end{document}